\documentclass[%
twocolumn,
nofootinbib,
 amsmath,amssymb,
 aps,
]{revtex4-2}

\usepackage{graphicx}
\usepackage{dcolumn}
\usepackage{bm}
\usepackage{hyperref}
\usepackage{xcolor}
\usepackage{scrextend}
\usepackage{gensymb}

\begin{document}

\preprint{APS/123-QED}

\title{Rock Neutron Backgrounds from FNAL Neutrino Beamlines\\in the $\nu$BDX-DRIFT Detector}

\author{D. Aristizabal Sierra}
 \email{daristizabal@ulg.ac.be}
 \affiliation{Universidad T\'ecnica
 Federico Santa Mar\'{i}a - Departamento de F\'{i}sica\\
 Casilla 110-V, Avda. Espa\~na 1680, Valpara\'{i}so, Chile}

 \author{J. L. Barrow}
 \email{jbarrow@fnal.gov}
 \affiliation{The Massachusetts Institute of Technology, Department of Physics, 77 Massachusetts Avenue, Building 4, Room 304, Cambridge, MA 02139, USA}
 \altaffiliation[Also at ]{Tel Aviv Univ.}
 \altaffiliation[formerly of ]{The Univ. of Tennessee}

\author{B. Dutta}
 \email{dutta@physics.tamu.edu}
 \affiliation{Mitchell Institute for Fundamental Physics and Astronomy, Department of Physics and Astronomy, Texas A\&M University, College Station, TX}
 
\author{D. Kim}
 \email{doojin.kim@tamu.edu}
 \affiliation{Mitchell Institute for Fundamental Physics and Astronomy, Department of Physics and Astronomy, Texas A\&M University, College Station, TX}

\author{D. Snowden-Ifft}
 \email{ifft@oxy.edu}
 \affiliation{Physics Department, Occidental College, 1600 Campus Rd., Los Angeles, CA 90041}

\author{L. Strigari}
 \email{strigari@tamu.edu}
 \affiliation{Mitchell Institute for Fundamental Physics and Astronomy, Department of Physics and Astronomy, Texas A\&M University, College Station, TX}

\author{M. H. Wood}
 \email{wood5@canisius.edu}
 \affiliation{Department of Quantitative Sciences, Canisius College, 2001 Main St., Buffalo, NY}

\date{\today}

\begin{abstract}

The $\nu$BDX-DRIFT collaboration seeks to detect low-energy nuclear recoils from CE$\nu$NS or BSM interactions at FNAL.  Backgrounds due to rock neutrons are an important concern. We present a~\texttt{GENIE} and~\texttt{GEANT4} based model to estimate backgrounds from rock neutrons produced in neutrino-nucleus interactions within the rock walls surrounding the underground halls. This model was bench-marked against the $2009$ COUPP experiment performed in the MINOS hall in the NuMI neutrino beam, and agreement is found between experimental results and the modeled result to within $30\%$.
Working from this validated model, a similar two-stage simulation was performed to estimate recoil backgrounds in the $\nu$BDX-DRIFT detector across several beamlines. In the first stage utilizing~\texttt{GEANT4}, neutrons were tallied exiting the walls of a rectangular underground hall utilizing four different neutrino beam configurations. These results are presented for use by other underground experiments requiring estimations of their rock neutron backgrounds. For $\nu$BDX-DRIFT, the second stage propagated neutrons from the walls 
and recorded energy deposited within a scintillator veto surrounding the detector and nuclear recoils within the detector's fiducial volume. The directional signal from the $\nu$BDX-DRIFT detector allows additional background subtraction. A sample calculation of a $10\,$m$^3\cdot\,$yr exposure to the NuMI Low Energy (LE) beam configuration shows a CE$\nu$NS signal-to-noise ratio of $\sim$2.5.

\end{abstract}

\maketitle

\newpage
\quad
\newpage
\quad
\newpage
\section{Introduction}

The $\nu$BDX-DRIFT detector is a directional time projection chamber (TPC) suitable for measurements of nuclear recoils produced by coherent elastic neutrino-nucleus scattering (CE$\nu$NS)~\citep{Freedman:1973yd,Freedman:1977xn} and by new physics interactions within the neutrino and dark-sectors, including those such as light (MeV) dark matter (DM)~\citep{PhysRevD.104.033004}. Its directional capabilities offer a unique environment for the identification of beyond Standard Model (BSM) signals~\citep{Abdullah:2020iiv}. The detector can operate with a variety of target nuclei, e.g. H, C, S and possibly Pb~\citep{PhysRevD.104.033004}. 
Studies of the performance of the detector using decay-in-flight neutrinos produced in the Long Baseline Neutrino Facility (LBNF) beamline at Fermi National Accelerator Laboratory (FNAL)~\citep{DUNE:2016evb} have been presented in Ref.~\citep{PhysRevD.104.033004}. These results have demonstrated that, with reasonable exposures ($10\,\text{m}^3$ for $7$ years of data taking), the detector will be able to measure $\sim300$-$400$ CE$\nu$NS events across various target materials. The resulting large statistics will in turn enable measurements of Standard Model (SM) electroweak and nuclear parameters, as well as searches for neutrino non-standard interactions (NSI), among others.

After the first measurements of CE$\nu$NS using CsI and liquid argon (LAr) detectors by the COHERENT collaboration~\citep{COHERENT:2017ipa,COHERENT:2020iec} at Oak Ridge National Laboratory's Spallation Neutron Source (ORNL SNS), an effort to undertake further measurements across other target nuclei
and different energy spectra utilizing various neutrino sources continues globally~\cite{Abdullah:2022zue}. Low energy experiments using reactor neutrinos are underway~\citep{CONNIE:2019swq,MINER:2016igy,Strauss:2017cuu,RED-100:2019rpf,Colaresi:2022obx,Billard:2016giu,CONUS:2020skt}, as well as further experiments at the ORNL SNS~\cite{Akimov:2022oyb}; this includes planning stages for the SNS Second Target Station, along with the European Spallation Source~\citep{Garoby:2017vew}. As a part of this global effort, the $\nu$BDX-DRIFT detector can provide a new and complementary avenue if it was to be based at FNAL: it would utilize decay-in-flight neutrinos and thereby observe higher energy regimes than the other global suite of experiments. Further, its technology offers measurements of the angular spectrum, in addition to the recoil energy spectrum; thus, in principle, cross section measurements in kinematic variables pertaining to the nuclear recoil
are possible.

However, the viability of all the above depends critically on background levels. Neutrino-induced neutrons produced in the rock, so-called ``rock neutrons", produce recoil-like backgrounds which are problematic and occupy the majority of discussions within this paper. The rock neutrons can be produced directly from neutrino-nucleus collisions, or when other neutrino-nucleus end-state particles interact in the surrounding material, generating still more neutrons. As will be shown below, rock neutrons produced in these ways have energies up to $\sim100\,$MeV and can produce nuclear recoils $\sim100\,$keV which themselves are expected from CE$\nu$NS and BSM interactions~\citep{PhysRevD.104.033004}. Recoils produced inside the shielding material around $\nu$BDX-DRIFT were considered in Ref.~\citep{PhysRevD.104.033004}, where it was shown that an expected signal-to-background ratio of better than $23$ could be achieved. Rock neutrons produced in the much larger volume of rock surrounding the underground facilities at FNAL are harder to estimate as the calculation must convolve the neutrino energy spectrum and interaction cross section on a variety of nuclei, the propagation of all end-state particles through the rock to the experimental hall, the possible interactions with shielding surrounding the detector, and, finally, the generation of nuclear recoils inside the fiducial volume of the detector.

The procedure presented here relies first upon a Monte Carlo neutrino event generator package,~\texttt{GENIE}~\citep{Andreopoulos:2009rq}, accounting for interactions of the neutrino beam with the rock material
in the surrounding walls of the FNAL underground MINOS experimental hall~\citep{MINOS:1998kez}. This first step is followed by a~\texttt{GEANT4}~\citep{GEANT4:2002zbu} simulation, which accounts for the propagation of the end-state particles generated in the~\texttt{GENIE} calculation and which potentially can enter the detector fiducial volume. The procedure is bench-marked with the aid of the COUPP beam-tagged data, which provides information on neutron-induced nuclear recoils. Four independent simulations will be presented based on four different neutrino flux configurations (NuMI LE and HE modes~\citep{Adamson:2015dkw} as well as DUNE on-axis and $39\,$m off-axis~\citep{DUNE:2020ypp}), and so collectively provide information not only valuable for a potential $\nu$BDX-DRIFT physics program but also for future neutrino detectors at FNAL. The results to be presented here can thus be understood as being aligned with and complementary to current efforts at the Accelerator Neutrino Neutron Interaction Experiment (ANNIE) at FNAL~\citep{ANNIE:2015inw}. Finally, results will be presented for rock neutron backgrounds in the fiducial volume of the $\nu$BDX-DRIFT with strong background protections afforded from the surrounding scintillator and the directionality of the interaction.

The remainder of this paper is organized as follows. In Sec.~\ref{sec:physics_capabilities} we provide a detailed discussion of the physics capabilities of the $\nu$BDX-DRIFT detector. In Sec.~\ref{sec:coupp}, details of the beam-tagged COUPP data are presented. In Sec.~\ref{sec:model}, the inputs used in the~\texttt{GENIE}-\texttt{GEANT4} Monte Carlo simulations are given. Results of the~\texttt{GENIE} output for final state particles are presented, along with the nuclear recoil spectrum in the COUPP detector's fiducial volume. In Sec.~\ref{sec:stage_I}, the neutron energy, zenith and azimuth spectra are provided for all four simulations, while in Sec.~\ref{sec:stage_II} these results will be used as input for the determination of the neutron background in the $\nu$BDX-DRIFT detector fiducial volume. Finally, in Sec.~\ref{sec:conclusions}, a summary and conclusions will be presented.

\section{Physics capabilities of the $\nu$BDX-DRIFT detector}
\label{sec:physics_capabilities}
Measurements of CE$\nu$NS within the $\nu$BDX-DRIFT detector will provide data enabling: (i) the determination of SM parameters, and (ii) searches for new interactions in the neutrino sector. These measurements can also enable searches for MeV-scale DM candidates produced in collisions of a proton beam on a fixed target. Detection proceeds by observation of the nuclear recoils produced by either of these progenitors within the fiducial volume of the detector.

Focusing on (i), the measurements which can be carried out include a precision determination of the weak mixing angle at $\sqrt{Q^2}\simeq 100\,$MeV, and the determination of the neutron root-mean-square (rms) radius of nuclides for which no data is yet available. As for (ii), searches include NSIs, interactions mediated by light vectors and scalars, along with sterile neutrinos. Analysis of these types of interactions have been completed using COHERENT and other reactor CE$\nu$NS data (see e.g.~\citep{Papoulias:2017qdn,AristizabalSierra:2019ufd,AristizabalSierra:2019ykk,AristizabalSierra:2022axl,Coloma:2019mbs}). Results from $\nu$BDX-DRIFT will thus prove complementary, while testing these hypotheses in a different energy domain and with different detector technologies.

As a function of detector operation pressure, CE$\nu$NS event rates in CS$_2$ peak at about $400\,$Torr. For a $10\,\text{m}^3$ detector operating over $7\,$years, the expected rate is on the order of $400\,$events. For CF$_4$ and utilizing the same operation pressure, the event yield increases by about a factor of two. With C$_8$H$_{20}$Pb, although with a lead target, the event yield is smaller because of the rapid loss of coherence. However, the statistics combined with the detector features are still large enough for the analysis of a few physics cases. Demanding isolation of lead-induced events, to study lead nuclear properties, fixes the operation pressure in that case to $\sim 5$ Torr~\citep{PhysRevD.104.033004}. 

Using CF$_4$ (C$_8$H$_{20}$Pb) as material target, a $10\,\text{m}^3$ detector operated at the pressures mentioned above will be able to measure the carbon and fluorine (lead) neutron rms with a $\sim3$\% ($\sim5$\%) precision. Ref.~\citep{PhysRevD.104.033004} has reported the following $1\sigma$ measurements
\begin{align}
    \label{eq:neutron_rms_C_S}
    r_\text{rms}^n|_\text{C}&=2.84^{+0.13}_{-0.15}\,\text{fm}\ ,
    \nonumber\\[1mm]
    r_\text{rms}^n|_\text{Pb}&=5.50^{+0.30}_{-0.29}\,\text{fm}\ .
\end{align}
Measurements for carbon and fluorine through electroweak neutral current processes do not exist, so these results provide valuable information for a better understanding of nuclear properties of light nuclide. For lead the result is not as competitive as that derived from PREX measurements~\citep{Abrahamyan:2012gp,Horowitz:2013wha}, but can be understood as complementary to it.

\begin{figure}
  \begin{centering}
  \centering
  \includegraphics[scale=0.16]{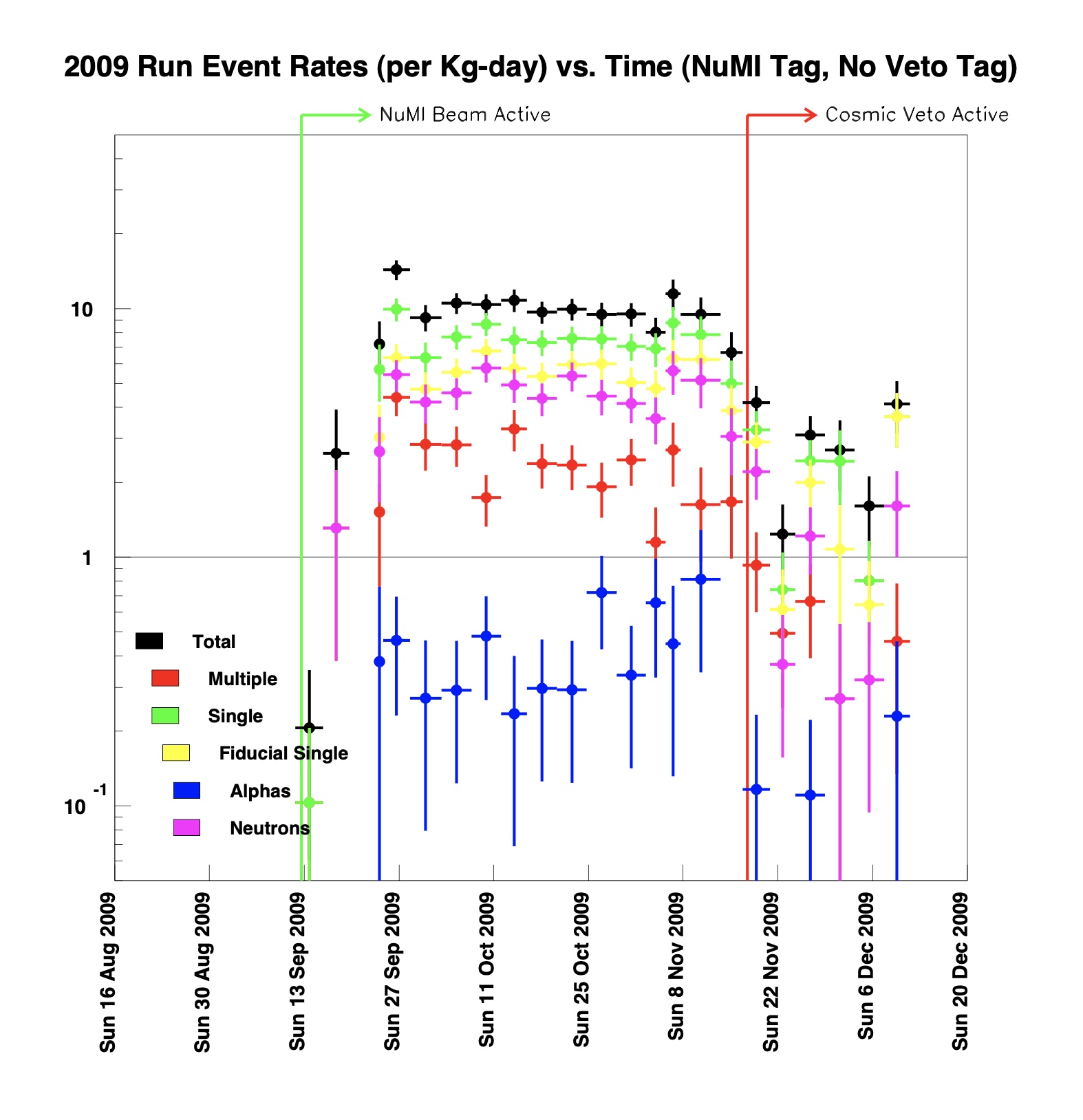}
  \caption{The 2009 COUPP bubble formation data tagged to the beam pulse. Published here with the permission of the COUPP collaboration.}
  \label{fig:COUPP}
  \end{centering}
\end{figure}
Studies of the weak mixing angle in CS$_2$ and CF$_4$ result in the following $1\sigma$ measurements
\begin{align}
    \label{eq:weak_mixing_angle}
    \sin^2\theta_W|_{\text{CS}_{2}}&=0.238 ^{+0.020}_{-0.016}\ ,
    \nonumber\\[1mm]
    \sin^2\theta_W|_{\text{CF}_{4}}&=0.238 ^{+0.021}_{-0.017}\ ,
\end{align}
both for $\sqrt{Q^2}\subset [78,397]\,$MeV, a renormalization scale for which at present no data is available. Interestingly enough, these results exceed what so far COHERENT measurements have achieved (see e.g.~\citep{Papoulias:2017qdn,Miranda:2020tif}) and are competitive with those expected from DUNE using the electron channel~\citep{deGouvea:2019wav}.

Searches for NSI in CS$_2$ can explore muon flavor related effective couplings. Sensitivities can improve by about a factor 2-3 upon current limits. To a certain extent they are not very sensitive to backgrounds (assuming reasonable amounts) nor to quark flavor. The $1\sigma$ measurements that can be achieved are given by \citep{PhysRevD.104.033004},
\begin{align}
    \label{eq:NSI}
    \epsilon_{\mu\mu}&=[-0.013,0.011]\oplus[0.30,0.32]\ ,
    \nonumber\\
    \epsilon_{e\mu}&=[-0.064,0.064]\ .
\end{align}
As has been emphasized, in order to achieve these goals a detailed understanding of rock neutron backgrounds becomes mandatory. The following sections focus on that.
\begin{table*}[t!]
    \centering
    \renewcommand*{\arraystretch}{1.2}
    \begin{tabular}{|c||c|c|c|c|c|c|c|c|c|}
     \hline
     \multicolumn{10}{|c|}{\textbf{Composition in rock at FNAL}}
     \\\hline\hline
     Isotope & $^{1}_{1}$H & $^{12}_{6}$C & $^{16}_{8}$O & $^{23}_{11}$Na & $^{27}_{13}$Al & $^{28}_{14}$Si
     & $^{39}_{19}$K & $^{40}_{20}$Ca & $^{56}_{26}$Fe\\\hline
     Composition [$\%$] & 1.5 & 1.1 & 56.4 & 0.3 & 9.5 & 24.2 & 0.9 & 4.3 & 1.8\\\hline
    \end{tabular}
    \vspace{0.2cm}
    
    \begin{tabular}{|c||c|c|c|}
        \hline
        \multicolumn{4}{|c|}{\textbf{Input parameters used in the simulations}}
        \\\hline\hline
        Beamline \& Mode & (POT/Pulse)$\times10^{13}$ & (Inter/Pulse/$\text{m}^3)\times10^{-4}$ & Period [s]\\
        \hline
        NuMI LE (c. 2009) & $2.88$ & $204.42$ & $2.43$ \\
        \hline
        NuMI LE & $4.00$ & $283.92$ & $1.3$ \\
        \hline
        NuMI HE & $4.00$ & $1277.69$ & $1.3$ \\
        \hline
        DUNE On-Axis at $1.2\,$MW & $7.5$ & $1142.23$ & $1.2$\\
        \hline
        DUNE $39\,$m Off-Axis at $1.2\,$MW & $7.5$ & $9.89$ & $1.2$\\
        \hline
    \end{tabular}
    \caption{\textbf{Upper:} The percentages of various nuclear isotopes in the rock, taken from discussions with FNAL experts.
    \textbf{Lower}: Summary of the input parameters for the models considered in this paper. The numbers of POT per pulse for NuMI and DUNE have been taken from Refs.~\citep{DUNE:2016evb,Adamson:2015dkw}.}
    \label{tab:rock_components}
\end{table*}

\section{COUPP}
\label{sec:coupp}

In order to present reliable results for nuclear recoil background predictions within the $\nu$BDX-DRIFT detector, any simulation used to predict such backgrounds requires bench-marking against data. Fortunately, such data exists. In 2009, the COUPP DM collaboration performed an experiment in the MINOS hall on-axis to an active NuMI beam~\citep{COUPP2011} at FNAL. COUPP was a bubble chamber experiment with a $15$-$20\, $keV threshold for detecting nuclear recoils filled with 3.5 kg of CF$_3$I~\citep{COUPP2011}. As discussed in~\citep{COUPP2011}, COUPP was a threshold detector providing no information on recoil energy or particle (nucleus) identification. Additionally, COUPP had no sensitivity to $\beta$, $\gamma$, or minimum ionizing particles. Using acoustic information $\alpha$ particle discrimination was possible~\citep{COUPP2011}.

In 2009, events were tagged as occurring when the beam was on or not. For the DM data analysis, only events uncorrelated with the beam were analyzed and published. However, unpublished, beam-tagged data from the COUPP collaboration was obtained~\citep{Priv_Comm}; a summary of these findings can be seen in Fig.~\ref{fig:COUPP}. The pink data points are single, fiducial events not tagged as $\alpha$ particles and are interpreted here as nuclear recoil events. The average of these data--taken from September 27, 2009 to November 8, 2009--is $4.65\pm0.19\,$events$/$kg$\cdot$day. During this running period, the cosmic veto was not operational; thus, some fraction of these events were caused by non-beam-related particles. To estimate this background, non-beam-related, background data taken during this time were averaged. Using a $100\,$ms timing window; the background rate due to random coincidences was estimated to be $0.0863\pm0.0074\,$events$/$kg$\cdot$day. Subtracting this from the observed rate gives, a true, beam-related nuclear recoil rate of $4.56\pm0.19\,$events$/$kg$\cdot$day to be compared to predictions.

\subsection{The Model}
\label{sec:model}

\begin{figure*}
  \begin{centering}
  \centering
    \includegraphics[scale=0.5]{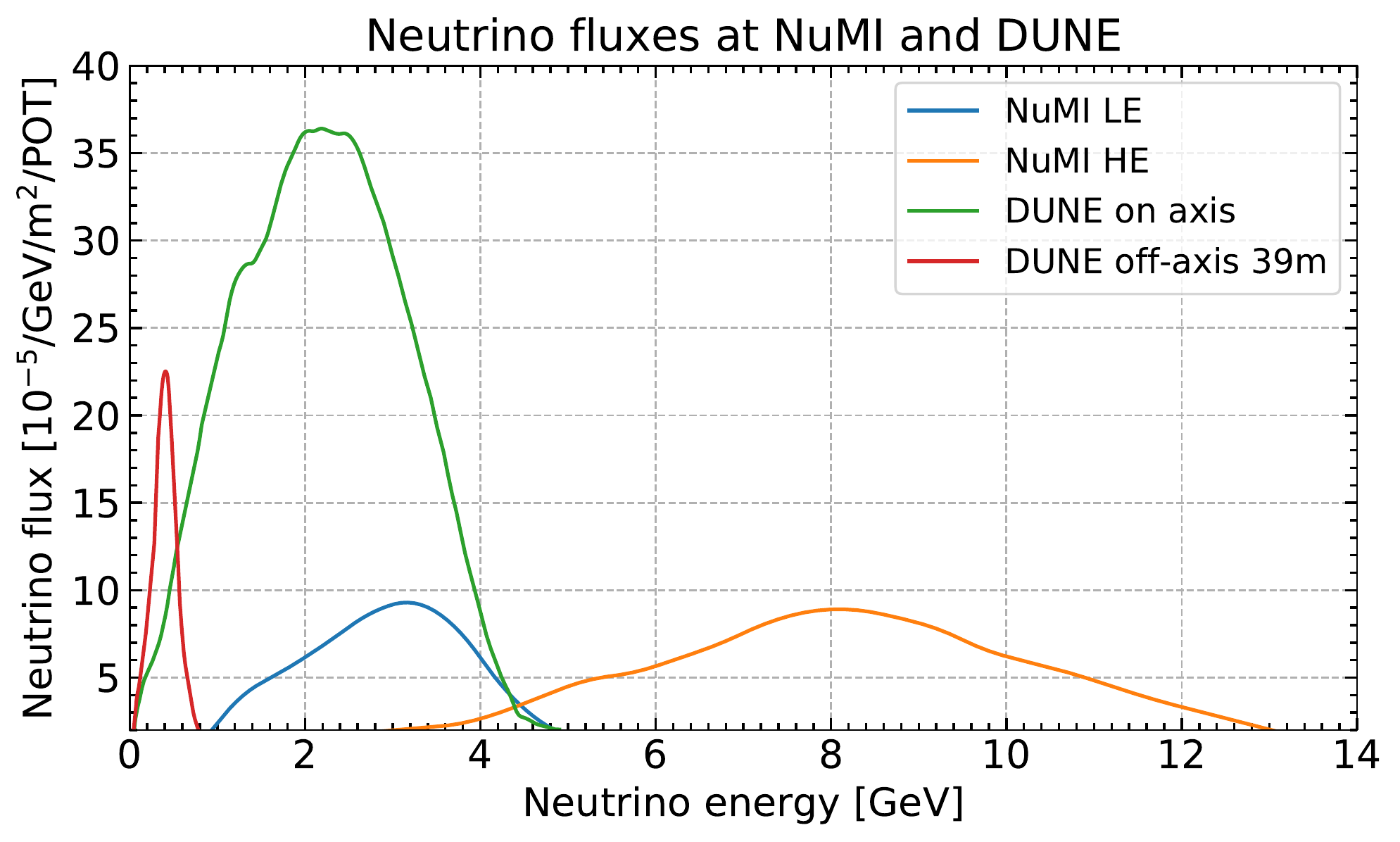}
    \caption{The $\nu_{\mu}$ energy spectra for various locations at FNAL: Fluxes at $1040\,$m downstream at NuMI in the LE and HE mode and at DUNE at $574\,$m downstream as well for the on-axis and the off-axis $39\,$m configurations. Results for NuMI are adapted from Ref.~\citep{Kopp:2007iq}, while for DUNE from the DUNE Technical Design Report (Fig. 4.9)~\citep{DUNE:2020ypp}. For the purposes of this study, small deviations in shape and rate between the $\nu_{\mu}$ and $\overline{\nu}_{\mu}$ horn modes spectra are ignored and are utilized identically.
  \label{fig:nuFlux}}
  \end{centering}
\end{figure*}

The parameters and model for backgrounds in the COUPP 2009 exposure to the neutrino beam are presented here. The composition of the rock can be seen in Table~\ref{tab:rock_components} (upper Table), and was assumed to be at a density of $2.33\,$g/cm$^3$. From the FNAL Data Logger~\citep{DataLogger}, the average number of protons on target (POT) per pulse was $2.88\times10^{13}$ with an average period of $2.43\,$s. These parameters as well as other assumed parameters are summarized in Table~\ref{tab:rock_components} (lower Table). The neutrino flux at the COUPP location was taken from~\citep{Kopp:2007iq} and increased by a factor of $(1040/939)^{2}$ due to the upstream location of the COUPP experiment relative to the originally assumed location~\citep{Kopp:2007iq}. Fig.~\ref{fig:nuFlux} shows the resultant flux, alongside several others to be discussed below.

According to MINOS logs~\citep{MinosLogs}, the NuMI beam was in reverse horn current mode during the COUPP 2009 run, implying predominately antineutrino production during the run period. Given the on-axis nature of the COUPP detector, it is expected that few differences exist between the $\nu_{\mu}$ and $\overline{\nu}_\mu$ fluxes (horn current settings) across the various NuMI beam energy settings~\citep{AliagaSoplin:2016shs}. Despite $\nu_{\mu}$ contamination of the $\overline{\nu}_\mu$ beam at high energies, we consider this single neutrino-type approximation robust, especially given the comparative lack of neutrons (which yield the most background events) entering the final state via charged current $\nu_{\mu}$ interactions.

\subsection{GENIE Event Generation}
\label{sec:GENIE_event_generation}
\begin{figure*}
    \centering
    \includegraphics[scale=0.5]{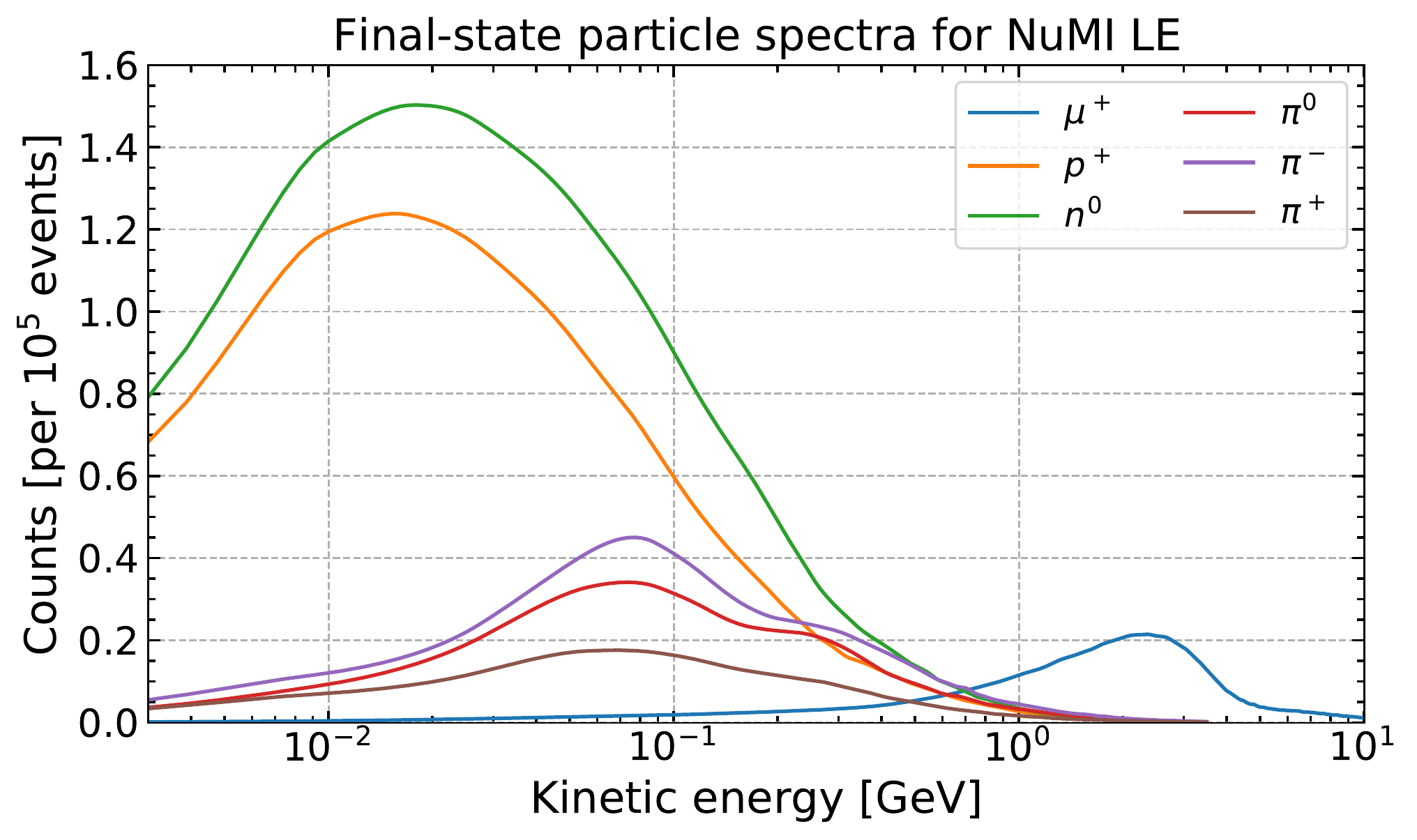}
    \caption{Energy spectra for $n$, $p$, $\pi^-$, $\pi^+$, $\mu^+$ and $\pi^0$     end-states of $\overline{\nu}_\mu$-nucleus interactions
    obtained by a~\texttt{GENIE} Monte Carlo simulation, for the NuMI LE neutrino flux. These spectra are used as input for the~\texttt{GEANT4} simulation of the COUPP result.}
    \label{fig:GENIE_histograms_NuMI_LE}
\end{figure*}

Given the previously discussed inputs, simulation of primary particle production via NuMI $\overline{\nu}_\mu$ interactions \textit{within} the rock surrounding the COUPP detector could be undertaken. Neutral and charged current processes across the whole range of energies of the NuMI flux resulting from $\overline{\nu}_\mu$ scattering were considered, providing predictions for final state neutrons, protons, charged and neutral pions, and antimuons. Fig.~\ref{fig:GENIE_histograms_NuMI_LE} shows energy distributions of the six different final state particles considered in this model for the NuMI LE neutrino flux employed in the COUPP simulation.

These primary particle production simulations were completed using the~\texttt{GENIE} Monte Carlo event generator~\citep{Andreopoulos:2009rq}, a staple within the FNAL neutrino community. The~\texttt{G18\_10a GENIE} tune~\citep{Andreopoulos:2015wxa} was used as a baseline, and cross section splines for all constituent elements were produced across the whole NuMI LE energy range. The chosen tune utilizes the hA2018 final state interaction (intranuclear cascade) model~\citep{Niewczas:2019fro,Golan:2012wx}, which uses a table-based method to predict full final states.  A similar simulation was undertaken using the hN2018 final state interaction model, which employs a fully stochastic intranuclear cascade and generally provides final state predictions with higher final state nucleon multiplicities. The mixture of elements making up the rock served as a direct input to~\texttt{GENIE} for event production, creating single samples; generally, the samples used throughout the studies discussed here were $\sim10^6$ events in size. Histograms with $\sim 50\,$MeV/c binning were constructed for the 6 most abundant final state particle types, $n$, $p$, $\pi^-$, $\pi^+$, $\mu^+$ and $\pi^0$.  As an example Figure~\ref{fig:GENIE_histograms_NuMI_LE} shows the energy distributions for these 6 end-state particles for the NuMI LE configuration and the hA \texttt{GENIE} model.  These distributions were use to as inputs for ~\texttt{GEANT4} \footnote{Correlated, event-by-event simulation of primary interaction products is indeed possible, and future work will utilize such techniques.}.

\begin{figure}[h!]
  \begin{centering}
  \centering
  \includegraphics[width=8.5cm]{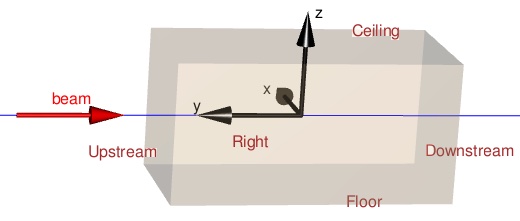}
  \includegraphics[width=8.5cm]{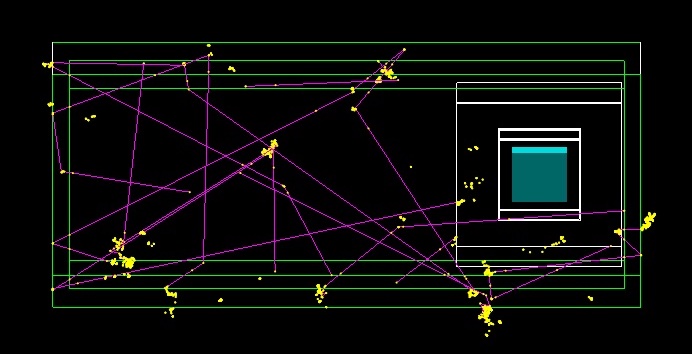}
  \caption{\textbf{Upper graph}: The labeled geometry of the underground experimental hall. \textbf{Lower graph}: A~\texttt{GEANT4} simulation  showing the location of the detector relative to the walls. The dimensions of the underground hall are 480/1070/427 cm in $x/y/z$. The aqua color shows the fiducial volume of the $\nu$BDX-DRIFT detector. The white frames show the location of the scintillator. Purple lines show neutrons trajectories. Yellow shows electron trajectories.
  \label{fig:HallFigure}}
  \end{centering}
\end{figure}
\subsection{GEANT4 Propagation}
\label{sec:GEANT4_propagation}

\begin{figure*}
  \begin{centering}
  \centering
  \includegraphics[scale=0.5]{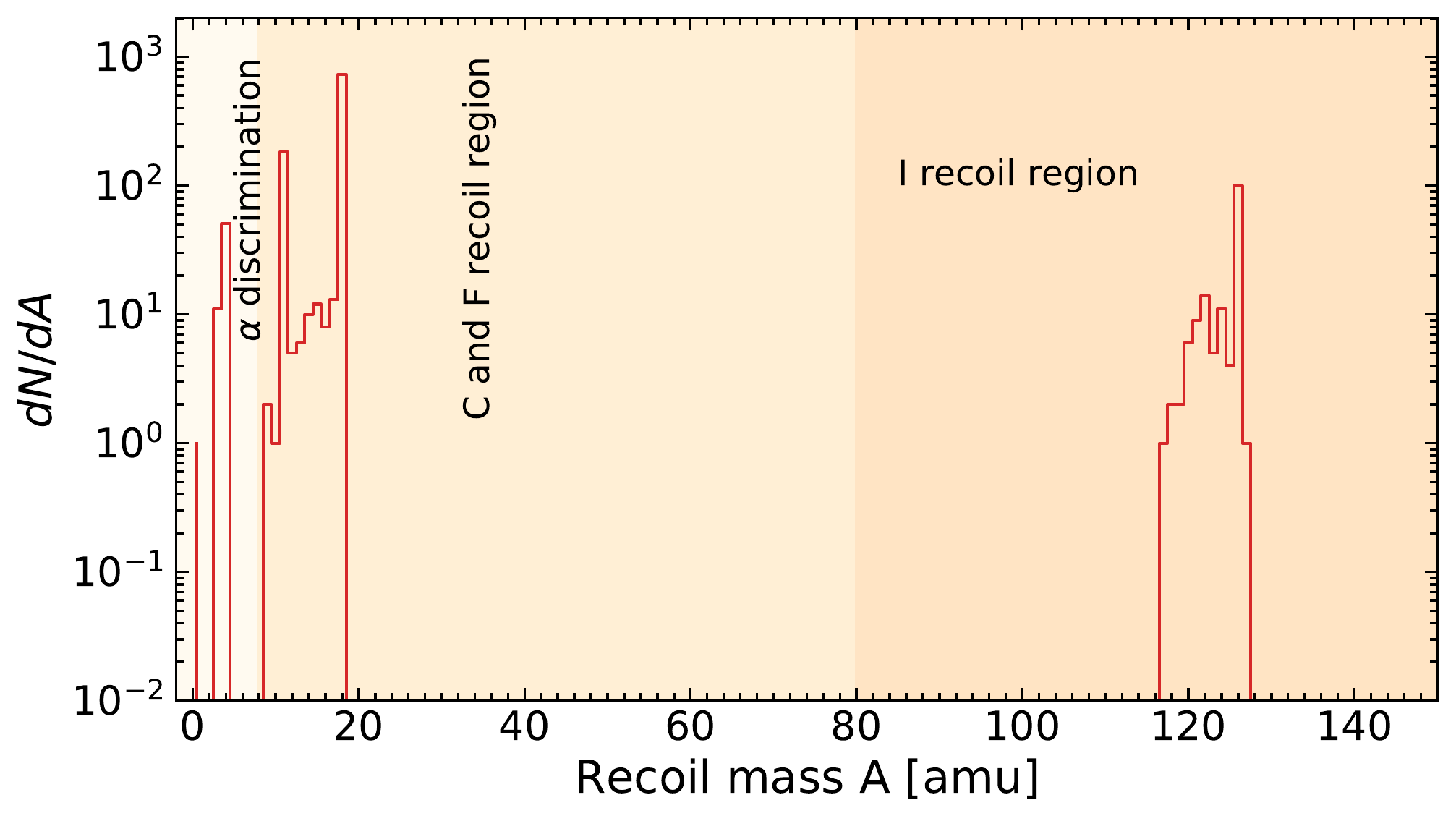}
  \caption{The spectrum of recoiling nuclei with kinetic energies ($E_r$) greater than 16.8 keV.  The small number of isotopes at masses other than C, F or I natural abundances are due to inelastic collisions between, mostly neutrons, and the target nuclei. Three regions in recoil masses are identified. Recoil masses in the region labelled ``$\alpha$ discrimination" were not counted because of $\alpha$ discrimination. Recoils in the region labelled ``C and F recoil region" were treated similarly, see text. Recoils in the third region, ``Iodide recoil region", were treated similarly as well. The shaded C, F and I regions are largely arbitrary, but there exist effectively no events within them beyond those at and slightly below the expected masses of these species. See text for further details.
  \label{fig:dN_dA}}
  \end{centering}
\end{figure*}

\texttt{GEANT4}~\citep{AGOSTINELLI2003250} was used to propagate the end-state particles from~\texttt{GENIE} through the rock and into the experimental hall and detector shown in Fig.~\ref{fig:HallFigure}. The dimensions of this hall (chosen to roughly approximate the size of the hallway where the COUPP experiment occurred) were considered small enough that uniform generation of end-state particles was assumed. The source considered in these simulations was taken as the rock walls, whose thickness was increased up to $2\,$m, at which point the observed rates in the detector stabilized. The COUPP detector was modeled as a cylindrical fiducial volume 15 cm in diameter and 12 cm high filled with CF$_3$I.  This was surrounded on almost all sides with propylene glycol (C$_3$H$_8$O$2$) the exception being a water filled region above the CF$_3$I.  The outer dimensions of the these elements were 30 cm in diameter and 44 cm high.  Again we thank members of the COUPP collaboration for providing this information~\citep{Priv_Comm}.
All massive nuclear recoils in the CF$_3$I were analyzed.  Fig.~\ref{fig:dN_dA} shows the resulting nuclear recoil spectrum in nuclear mass.

The nucleation efficiency for bubble formation following nuclear recoil within the COUPP detector is given~\citep{BubbleNucleation} as,
\begin{equation}
    \label{eq:efficiency_COUPP}
    \epsilon(E)=1-e^{-\alpha[(E - E_T)/E_T)]}\quad (E > E_T)\ ,
\end{equation}

\noindent where $E_T$ is a universal threshold while $\alpha$ depends on the recoil type; $\alpha_\text{CF}$ (for Carbon and Fluorine recoils) was determined to be $0.15$ from AmBe neutron exposures, while $\alpha_\text{I} = 2.8^{+1.6}_{-0.8}$ (for Iodide recoils) and $E_T = 16.8^{+0.8}_{-1.1}\,\text{keV}$ were determined using a $12\,$GeV $\pi^-$ beam~\citep{BubbleNucleation}. For this work, the mean values of these quantities were employed; note that no uncertainty was given for $\alpha_{CF}$.

\texttt{GEANT4} events in which multiple bubbles were removed as the COUPP data reports only single events in the fiducial volume. \texttt{GENIE's} input simulation to~\texttt{GEANT4} utilizing the hA2018 model yields a predicted rate of $2.930\pm0.039\,$events$/$kg$\cdot$day. As a check on the effect of the geometry of the experimental hall on this result the length of the experimental hall was increased by a factor of 3. The result was $2.890\pm0.046\,$events$/$kg$\cdot$day in agreement with the previous result. For clarity these results, and the ones discussed below, are summarized in Table~\ref{tab:Rates}. The~\texttt{GENIE} hN model yields a rate of $3.081\pm0.025\,$events$/$kg$\cdot$day. 
These were averaged together to produce a predicted rate of $3.006\pm0.023\,$events$/$kg$\cdot$day. These events were created by, largely, rock neutrons entering the COUPP detector from the walls, thus creating recoils  which nucleated a bubble. 

Recoils can of course also be created directly inside the COUPP fiducial volume by direct neutrino scatters, the dominant component of these being non-CE$\nu$NS events such as neutrino-nucleon quasi-elastic scattering, a subdominant contribution from neutrino-nucleon scattering, resonant single pion production and by products of deep-inelastic scattering. To better understand this,~\texttt{GENIE} was run with CF$_3$I, instead of rock, as the target, and an overall rate for such scatters was $0.35\,$events$/$kg$\cdot$day. However, this total event count ignores the fact that not all such events will nucleate a bubble. For some events, no large remnant nuclei survive; for those that do survive, there is a less than 100\% chance of nucleating a bubble given their momentum\footnote{Note that~\texttt{GENIE} is currently unable to record all the properties of remnant nuclei; similarly, for all but one nucleus (Oxygen), no photonic de-excitation occurs. There is motion within the community to include more of this necessary microphysics~\citep{Gardiner:2020ulp,Gardiner:2021qfr}, and we look forward to more updates to such tools.}. We therefore bracket our modeled results as $(3.006,3.356)\pm0.023\,$events$/$kg$\cdot$day. These event rates are to be compared to the experimental rate of $4.56\pm0.19\,$events$/$kg$\cdot$day. 

The predicted rate of this study sits roughly $30\%$ lower than the observed experimental rate. There are, however, a large number of systematics which could explain this difference. The bubble formation model has systematics associated with the assumptions discussed above, though these appear to be relatively small. For instance, varying the bubble formation parameters such as $\alpha_I$ and $E_T$ gives a $0.09\,$events/kg$\cdot$day systematic variation to the rate. \texttt{GENIE} and \texttt{GEANT} have systematics associated with the particular models chosen, and are largely unknown to this study without the use of a universe style approach. Slight changes in the geometric configuration of the detector can also contribute to the uncertainty. Similarly, the neutrino flux model is known to have large normalization uncertainties which have not been considered for this study.

\begin{table*}[t!]
    \centering
    \renewcommand*{\arraystretch}{1.2}
    \begin{tabular}{|c||c|}
        \hline
        \multicolumn{2}{|c|}{\textbf{Rate Comparison Summary}}
        \\\hline\hline
        \textbf{Source} & \textbf{Rate [events/kg$\cdot$day]} \\
        \hline
        \texttt{GENIE} hA & $2.930\pm0.039$ \\
        \hline
        \texttt{GENIE} hA w/3$\times$longer exp. hall & $2.890\pm0.046$ \\
        \hline
        \texttt{GENIE} hN & $3.081\pm0.025$ \\
        \hline
        \texttt{GENIE} hA, hN average & $3.006\pm0.023$ \\
        \hline
        Unshielded \textit{in-situ} & $0$ to $0.35$ \\
        \hline  \hline
        Prediction & $(3.006, 3.356)\pm0.023$ \\
        \hline
        Experiment & $4.56\pm0.19$\\
        \hline
    \end{tabular}
    \caption{This table summarizes the rates from various sources and, at the end, the final prediction range in comparison with the COUPP data.}
    \label{tab:Rates}
\end{table*}

\begin{table*}[t!]
    \centering
    \renewcommand*{\arraystretch}{1.2}
    \begin{tabular}{|c||c|c|c|}
        \hline
        \multicolumn{4}{|c|}{\textbf{Number of simulated particles}}
        \\\hline\hline
        Beamline \& Mode & Stage I [$\times$10$^6$] & Walls [$\times$10$^6$] & Stage II [$\times$10$^9$]\\
        \hline
        NuMI LE & 207 & 17.2 & 2.36 \\
        \hline
        NuMI HE & 130 & 2.66 & 1.70 \\
        \hline
        DUNE On-Axis at $1.2\,$MW & 434 & 8.26 & 2.36 \\
        \hline
        DUNE $39\,$m Off-Axis at $1.2\,$MW & 1660 & 5.51 & 2.10 \\
        \hline
    \end{tabular}
    \caption{Output table number of particles simulated at various stages. Column 2 shows the number of end-state particles simulated in Stage I (see Sec.~\ref{sec:stage_I}). Column 3 shows the number of neutrons entering the experimental hall from the walls. These neutrons were used to generate the distributions for the Stage II simulations (see Sec.~\ref{sec:stage_II}). Column 4 shows the number of neutrons simulated in Stage II restarted on the walls of the experimental hall. (see Sec.~\ref{sec:stage_II}).}
    \label{tab:model_numbers}
\end{table*}

\section{Stage I: Rock Neutron Results}
\label{sec:stage_I}

With the bench-marked model in hand we now turn to predicting backgrounds in future, planned experiments. As the COUPP results show, backgrounds due to rock neutrons in an unshielded detector are high, too high to accomplish the goals of the $\nu$BDX-DRIFT collaboration. We therefore include a scintillating veto around the simulated $\nu$BDX-DRIFT detector. The COUPP collaboration installed a scintillating veto around most of their detector with a resulting drop in un-vetoed rate after the period of unshielded running described above and shown in Figure~\ref{fig:COUPP}. That the rate did not drop further was the result of lack of shielding around the bottom of the detector; the shielding was designed to veto cosmic-ray generated events not beam events. For purposes of simulation we will assume the $\nu$BDX-DRIFT detector is surrounded by 75 cm of BC-521 organic scintillator on all sides, similar to the veto COUPP utilized. As will be shown below use of this veto drastically reduces the rate of events in the $\nu$BDX-DRIFT detector.

But as a result the simple, single-stage simulation used for the COUPP background calculation is impractical. A two-stage strategy was therefore adopted in which neutrons were recorded exiting the walls of the experimental hall. The hall was assumed to have an upstream and downstream wall perpendicular to the neutrino beamline and 4 walls parallel to the beamline as shown in Fig.~\ref{fig:HallFigure}. For each wall the energy and angular distributions of neutrons exiting the walls $\emph{for the first time}$ were recorded and smoothed. In a second stage, neutrons were restarted at the walls with the same energy and angular distributions with a resulting increase in simulation speed of roughly two orders of magnitude. The computed energy and angular distributions for all simulations are shown below for use in other applications.

To bracket the range of possibilities at FNAL four simulations were done. Table~\ref{tab:rock_components} (lower) summarizes the main input parameters for these simulations. The neutrino energy spectra for all simulations are shown in Fig.~\ref{fig:nuFlux}. All simulations assumed that the horn currents were set to predominantly produce $\overline{\nu}_\mu$s. $\overline{\nu}_\mu$s produce more neutrons than $\nu_\mu$s due to the nature of the charge current interaction, and, in terms of background, therefore represent a worst case scenario. The location of the COUPP detector was on the far upstream end of the MINOS hall, 939 m from the target. All NuMI simulations were done at this location. As before the fluxes for NuMI, from~\citep{Kopp:2007iq} assuming 1040 m from target, were increased by (1040/939)$^2$ to correct for this assumption. For the DUNE simulations the experimental hall, shown in Fig.~\ref{fig:HallFigure}, was located 574 m from the DUNE target at the location of the DUNE near detector hall. Two positions were chosen, on-axis and 39 m off-axis, to bracket the possibilities there. As shown in Fig.~\ref{fig:nuFlux} these positions have very different fluxes and energy spectra. Note that for the DUNE simulations it was assumed that the experimental hall shown in Fig.~\ref{fig:HallFigure} was completely surrounded by rock which is not what is planned for the near detector hall. The DUNE simulations, therefore, are more indicative of backgrounds generated on either side of the DUNE near detector hall. The total number of neutrino interactions per m$^3$ of rock per pulse is shown in column 3 of Table~\ref{tab:rock_components} (lower) for each beam and mode. 

\begin{figure*}[t!]
  \centering
  \includegraphics[scale=0.43]{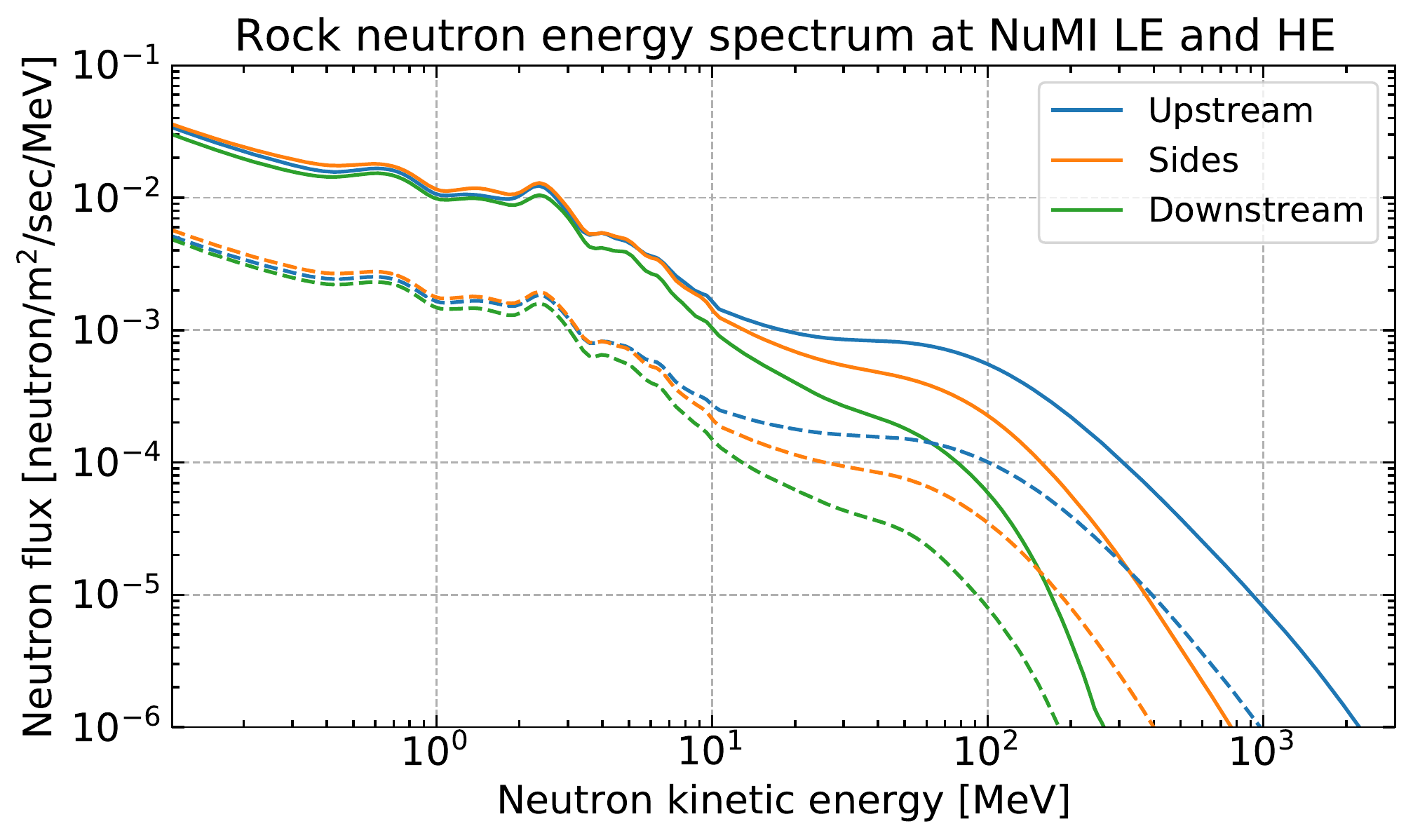}
  \includegraphics[scale=0.43]{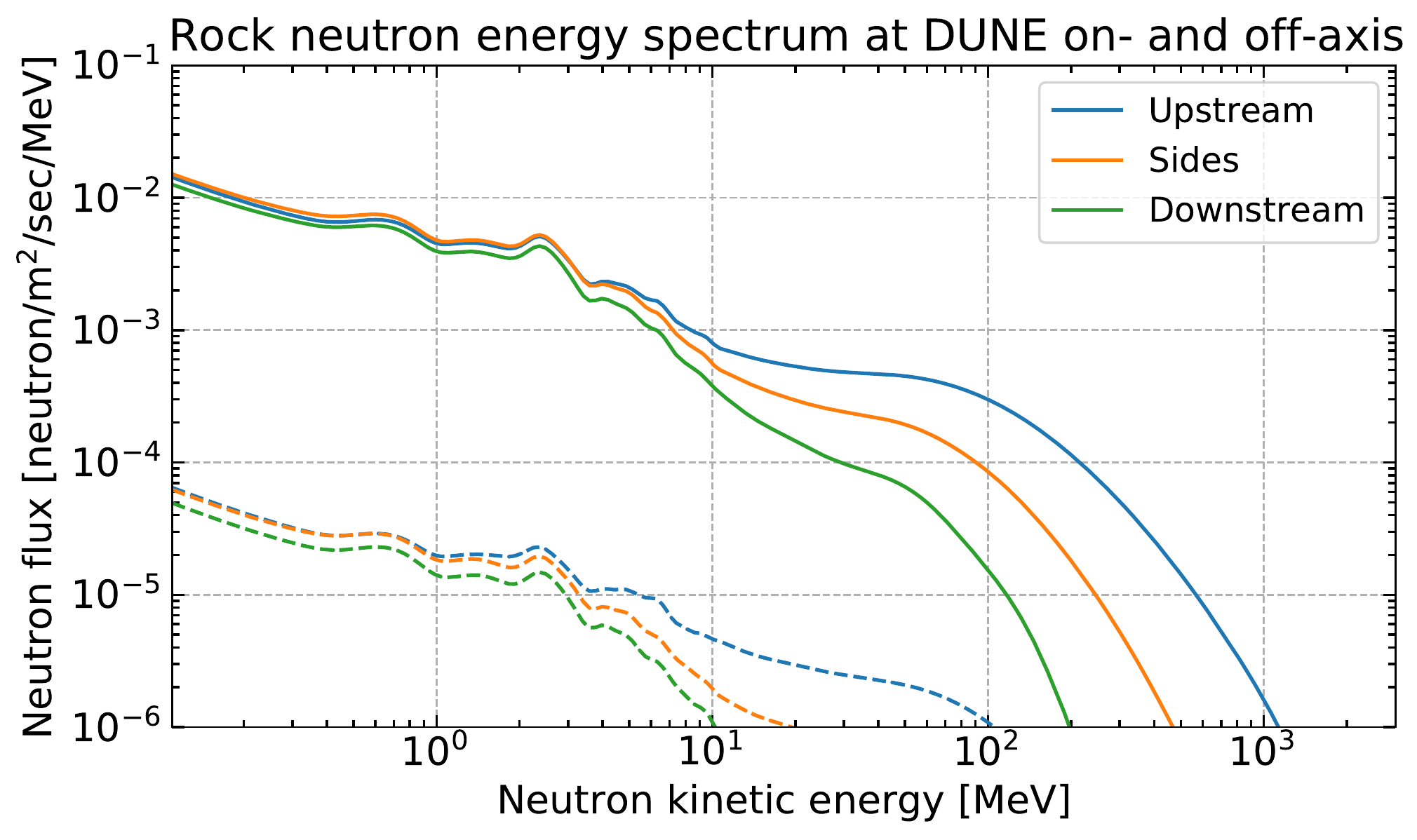}
  \caption{The energy distribution of rock-neutrons generated in the 4 simulations of Table~\ref{tab:rock_components}. The blue lines show the spectra coming from the upstream wall. The orange lines show the spectra coming from the side walls. And the green lines show the spectra coming from the downstream wall. In the left graph, the solid (dashed) curves correspond to results obtained with the NuMI HE (LE) neutrino mode. In the right graph---instead---to results derived with the DUNE on-axis (off-axis)} configuration.
  \label{fig:rock_neutron_spectra}
\end{figure*}
\begin{table*}[t]
    \centering
    \renewcommand*{\arraystretch}{1.2}
    \begin{tabular}{|c||c|c|c|c|}
        \hline
        \multicolumn{5}{|c|}{\textbf{Simulations output for neutron flux from the walls}}
        \\\hline\hline
        Beamline \& Mode & Upstream [$n^0$/s/m$^2$] & Sides [$n^0$/s/m$^2$]
        & Downstream [$n^0$/s/m$^2$] & Background [events/m$^3$/year]\\\hline
        NuMI LE & 0.0355 & 0.0204 & 0.0110 & 8.61 $\pm$ 0.62 \\
        \hline
        NuMI HE & 0.209 & 0.131 & 0.0727 & 54.9 $\pm$ 3.8 \\
        \hline
        DUNE On-Axis at $1.2\,$MW & 0.101 & 0.0276 & 0.0524 & 23.3 $\pm$ 1.3\\
        \hline
        DUNE $39\,$m Off-Axis at $1.2\,$MW & 0.000381 & 0.0000831 & 0.000162 & 0.0396 $\pm$ 0.0031 \\
        \hline
    \end{tabular}
    \caption{Output table shows neutron flux from different walls and background in the signal region. For details see Sec.~\ref{sec:stage_I}.}
    \label{tab:model_output}
\end{table*}

\begin{figure*}[t]
  \centering
  \includegraphics[scale=0.43]{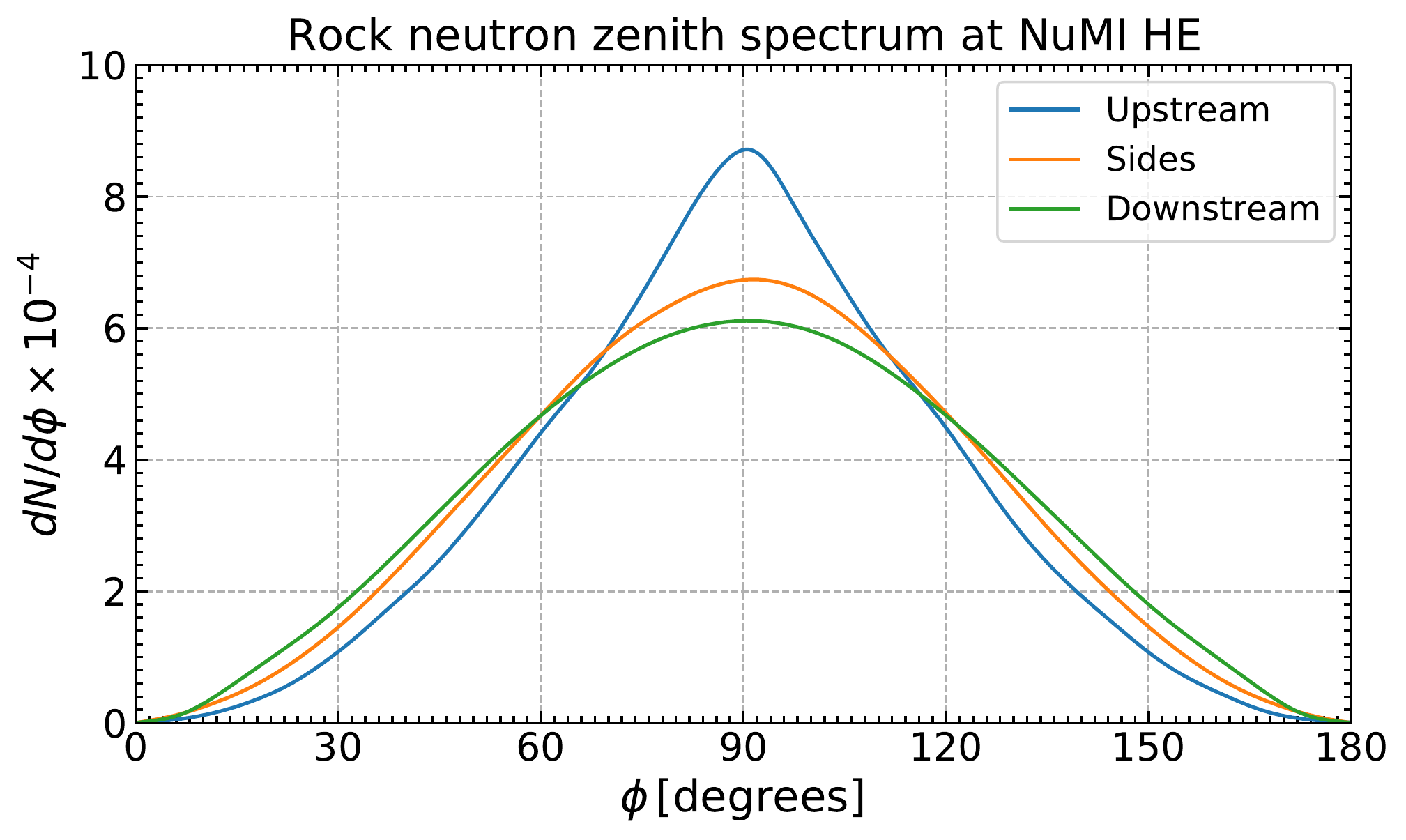}
  \includegraphics[scale=0.43]{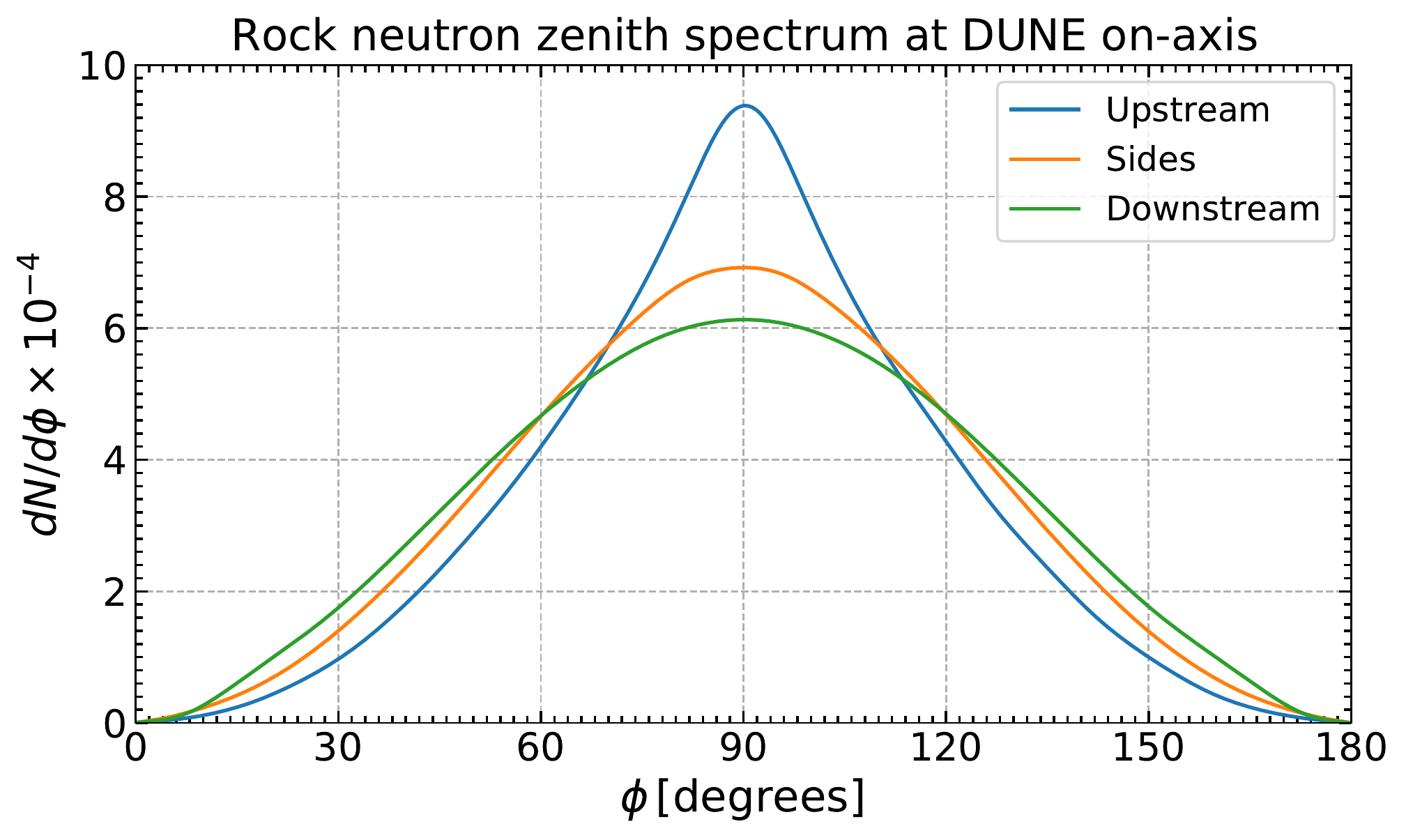}
  \includegraphics[scale=0.43]{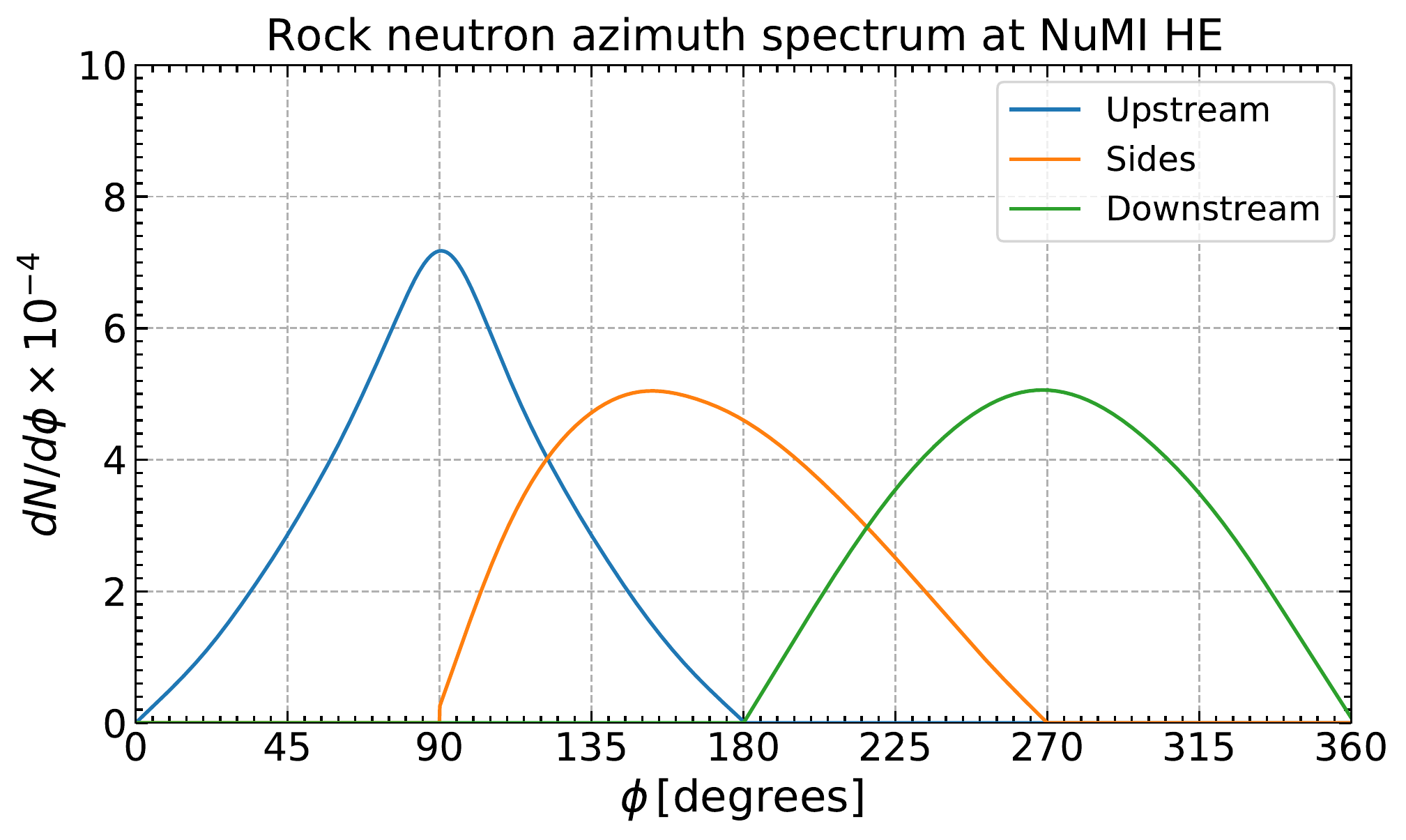}
  \includegraphics[scale=0.43]{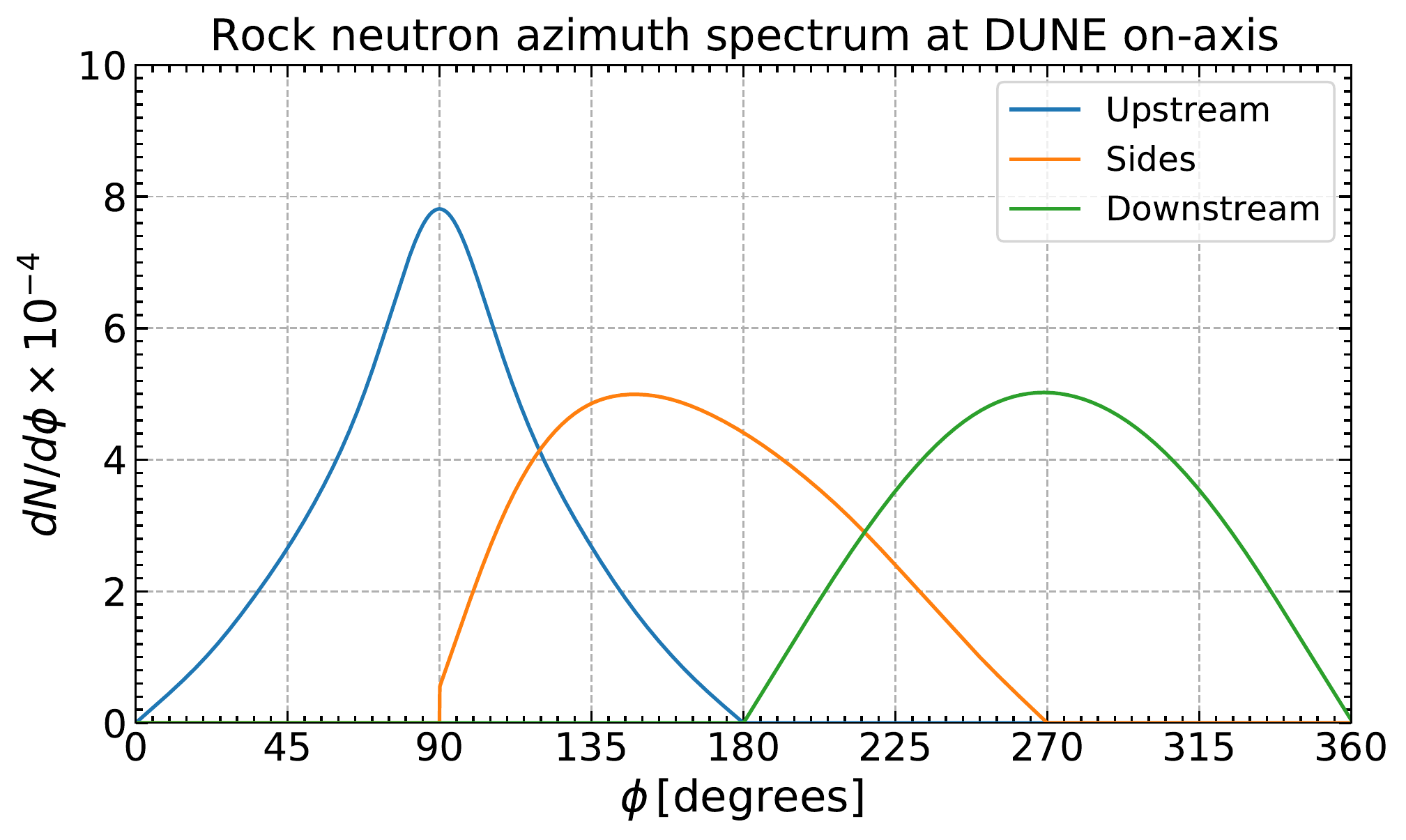}
  \caption{\textbf{Upper row graphs}: The zenith angle distribution of rock-neutrons generated in the NuMI HE (left graph) and the DUNE on-axis (right graph) simulations of Table~\ref{tab:rock_components}. The blue lines show the spectra coming from the upstream wall. The yellow lines show the spectra coming from the side walls. And the green lines show the spectra coming from the downstream wall. With rather small variations, results for the NuMI LE (DUNE off-axis) resemble those of the NuMI HE (DUNE on-axis) as so are not displayed. \textbf{Lower row graphs}: Same as for those on top, but for azimuth angle distribution. Results are presented for the same simulations as we have found that differences as well with the other two are negligible.
 \label{fig:angular_spectra}}
\end{figure*}

End-state particles from these interactions were propagated, using~\texttt{GEANT4}, to the walls of the experimental hall where, as discussed above, neutron characteristics were recorded and saved. Charged particles exiting the walls of the experimental hall were not saved as they would either range out in the scintillator or be vetoed there. Table~\ref{tab:model_numbers} shows the number of particles simulated at each stage of the simulation. The smoothed, rock-neutron energy distributions for the four simulations are shown in Fig.~\ref{fig:rock_neutron_spectra}. As expected the flux of neutrons exiting the walls is higher on the upstream wall than the downstream wall with a harder spectrum. The sides fall somewhere in between. Also for the same POT/pulse, see Table~\ref{tab:rock_components}, higher energy configurations produce higher fluxes of rock-neutrons. Table~\ref{tab:model_output} shows a summary of the output from the simulations. Columns 2, 3 and 4 show the rates for various surfaces relative to the beam. These numbers are nothing more than the integral of the differential flux, see Fig.~\ref{fig:rock_neutron_spectra}, with energy but they provide a simple way of comparing the various beamlines and modes.

Fig.~\ref{fig:angular_spectra} (upper row) shows the spectra of zenith angles, measured from the $z$-axis, for each of the walls. As expected the upstream wall shows a more pronounced peak than does the downstream wall. Results are shown only for the NuMI HE mode and the DUNE on-axis configuration. Results for the NuMI LE (DUNE off-axis 39 m) resemble rather closely those of the NuMI HE mode (DUNE on-axis) and so are not displayed. Fig.~\ref{fig:angular_spectra} (lower row) shows as well the spectra of azimuth angles, measured from the $x$-axis, for each of the walls. The zenith and azimuth angle specifies a vector which, adopting the~\texttt{GEANT} convention, points in a direction from which the particle came. The upstream wall therefore emits particles with azimuth angles from 0 to $\pi$, vectors which point into the rock, while the downstream wall emits particles from $\pi$ to $2\pi$, vectors which point into the experimental hall. Once again the upstream wall exhibits a more concentrated distribution as the emission of neutrons from the downstream wall would entail multiple bounces before emission from the wall. Finally the sides, right hand wall shown here, shows an asymmetric distribution skewed towards smaller azimuth angles indicating a preference for emission from the beam direction. 
In summary the angular distributions show a preference for neutron emission from the direction to the target which decreases from the upstream wall to the sides to the downstream wall.

\begin{figure*}[t]
  \centering
  \includegraphics[scale=0.42]{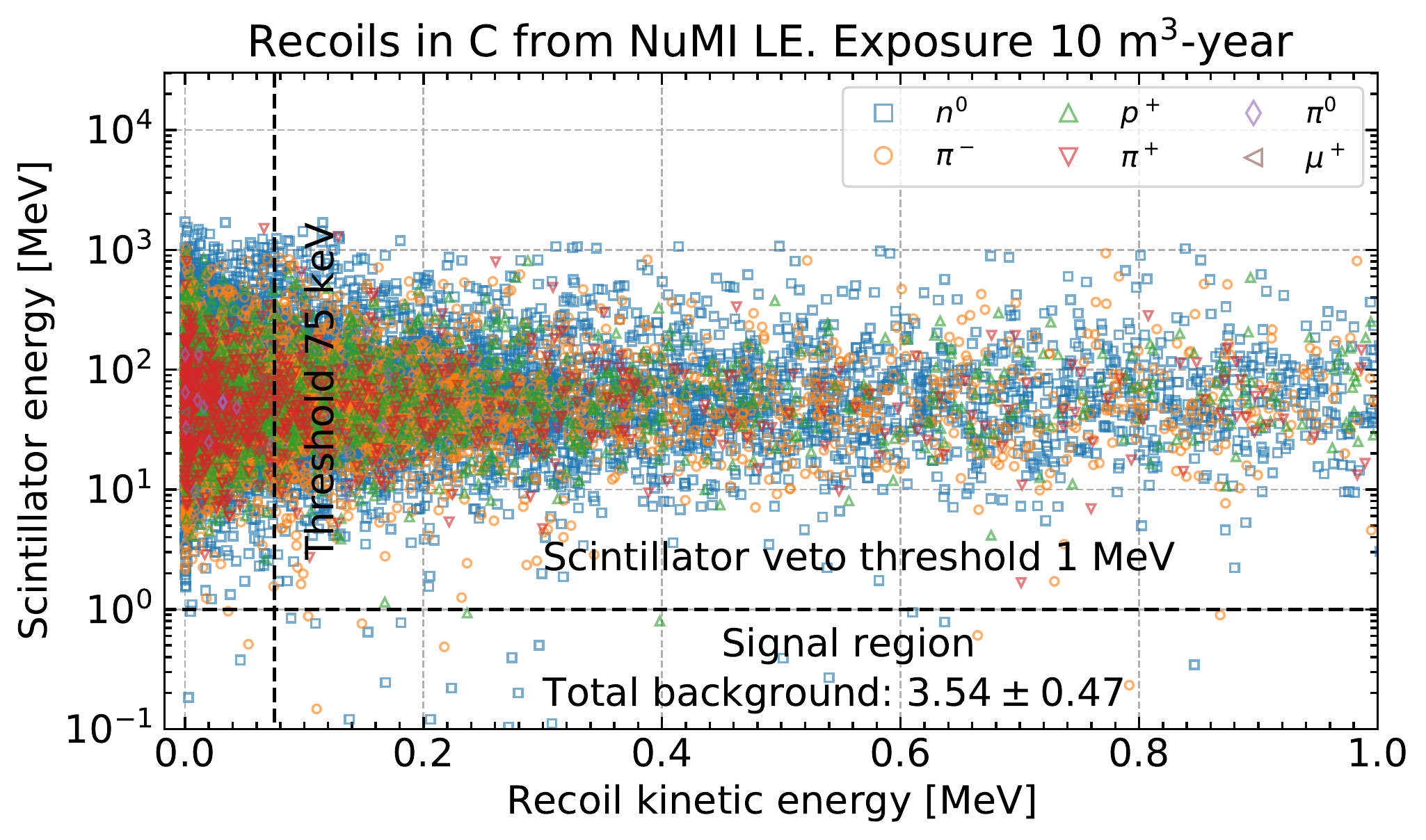}
  \includegraphics[scale=0.42]{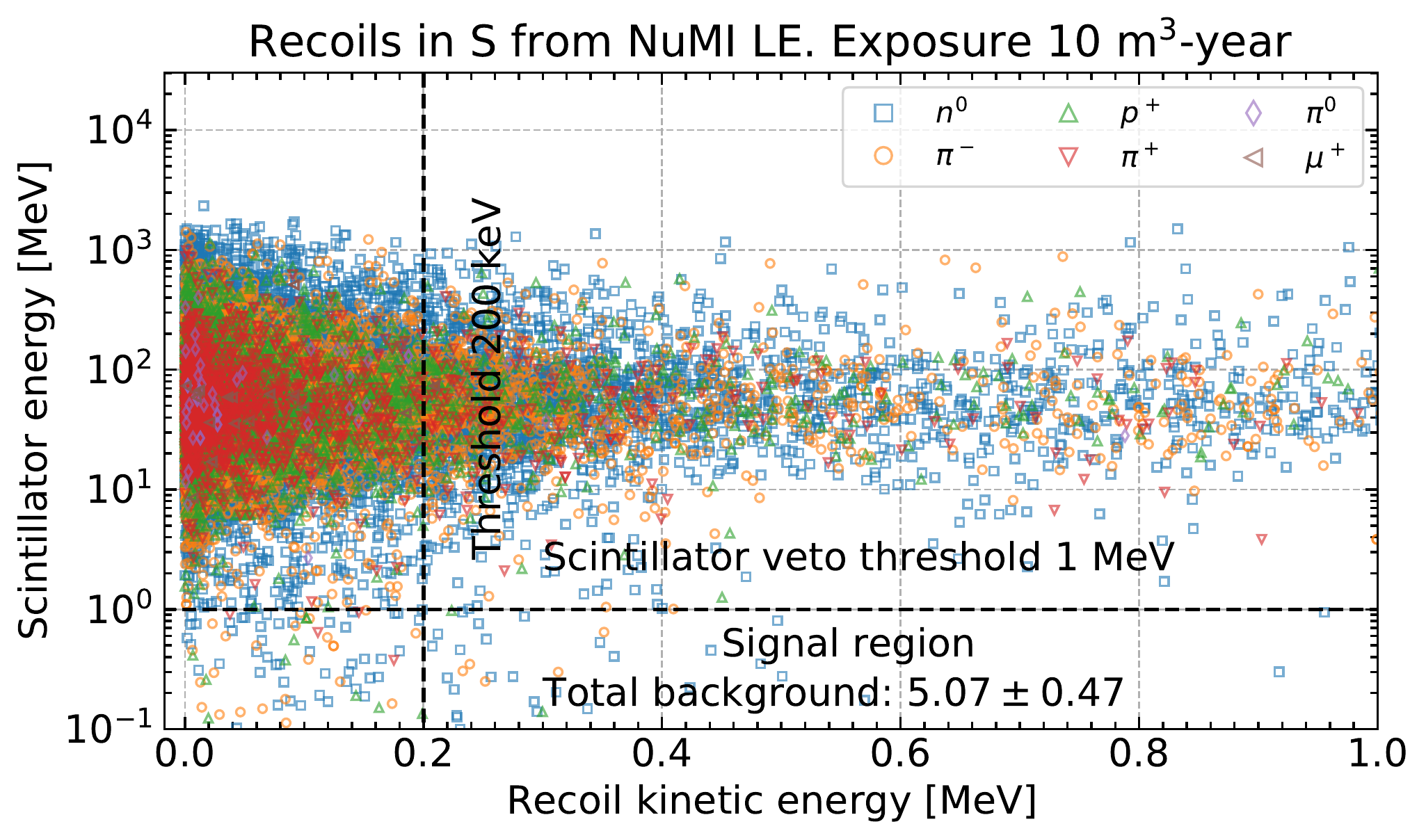}
  \caption{Plots showing the distribution of recoil energies vs energy deposited in the scintillator with the neutrino-induced end-state particle responsible for the recoil shown in different colors. The left graph shows the results for C recoils while the right graph shows the S recoils. Both are heavily dominated by neutron end-states (about 63\% for both target nuclei). The vertical dashed black lines indicate the recoil thresholds, 75 keV for C and 200 keV for S. The horizontal dashed black lines show the threshold for the scintillator veto, 1 MeV; events with larger energies are vetoed. The lower right region therefore shows the signal region where either CE$\nu$NS or BSM recoils events would occur. The background rate, 
  in events per m$^3$ per year, are shown there. The sum is shown in the fifth column of Table~\ref{tab:model_output}.}
  \label{fig:LE_time_normalized}
\end{figure*}
\section{Stage II: $\nu$BDX-DRIFT Results}
\label{sec:stage_II}
As discussed above the main motivation for this work is the reliable prediction of backgrounds for the $\nu$BDX-DRIFT experiment. To that end a Stage II simulation was set up and run to predict backgrounds. Neutrons were fired from the walls of the experimental hall with energy and angular spectra such as shown in Figures~\ref{fig:rock_neutron_spectra} and \ref{fig:angular_spectra}. 
From the outside in, the detector consisted of a $75\,$cm thick BC-521 scintillator veto surrounding the entire detector with outer dimensions of $3\,$m, a $0.5\,$inch thick stainless-steel, cubic vacuum vessel with outer dimensions of $1.5\,$m and a cubic fiducial volume for recoils composed of CS$_2$ at a density 2.44 times higher than $400\,$Torr. This increased pressure increases the efficiency for recording recoils while minimizing double recoils~\cite{AstroPle.28.4.409}; final results are corrected at the end.

\texttt{GEANT} recorded any energy deposited in the scintillator veto and in the fiducial volume. Fig.~\ref{fig:LE_time_normalized} shows the results for the NuMI LE beamline and mode. On the horizontal axis is the recoil kinetic energy for C and S. On the vertical axis is the amount of energy deposited in the scintillator. The different colors represent the end-state particles from $\overline{\nu}$-nucleus interactions which produced neutrons which entered the experimental hall and created C or S recoils in the fiducial volume of the $\nu$BDX-DRIFT detector. Neutron end-state particles from $\overline{\nu}$-nucleus interactions dominate the recoil rate. The vertical dashed line shows the kinetic energy threshold for recoil detection after~\citep{PhysRevD.104.033004}. As can be seen in these graphs a huge number of recoils are predicted above threshold. However the vast majority of nuclear recoils above threshold also come with an enormous deposition of energy in the scintillator, on order $100\,$MeV \footnote{It should be noted that the benchmarked COUPP 2009 experiment was mostly sensitive to 1-10 MeV neutrons while $\nu$BDX-DRIFT is mostly sensitive to 10-100 MeV neutrons due to the necessity of penetrating the scintillator.}. These large energy depositions occur due to showers produced as the neutrons traverse the detector and resulting charged particle interactions in the scintillator veto. The horizontal dashed line 
indicates a $1\,$MeV threshold on the veto; events with energy greater than this are vetoed. Signal events, CE$\nu$NS events or BSM interactions, would appear in the lower right corner of these graphs. Backgrounds, in this context, means events due to beam neutrons appearing in this lower right corner. The rate of recoils, and errors, for C and S appear in this lower right corner in Fig.~\ref{fig:LE_time_normalized} in units of events per m$^3$ per year. The fifth column in Table~\ref{tab:model_output} shows the background rates for each of the beamlines and modes studied in this paper. As can be seen the highest backgrounds occur in the NuMI beamline in the HE mode.

As a check of the Stage I of the simulation for this high background configuration, a run~\emph{with the scintillating veto in place} was completed. After firing $2.3\times10^9$ end-state neutrons from the walls the result was in statistical agreement with the Stage II neutron results to within $15\%$ validating the use of the multi-stage procedure.

There remains a question as to how these beam-related backgrounds compare to their non-beam-related cousins.  While this question has not been studied in detail, an \textit{indication} can be found when again considering the 2009 COUPP results~\citep{COUPP2011}. The COUPP collaboration found a neutron background of $3$ events across a $28.1\,$kg$\cdot$day exposure for a rate of about $0.1\,$events/kg$\cdot$day; this rate was measured with lower thresholds and while maintaining a scintillating shield similar to that described in this work. However, this rate was \textit{not} in coincidence with the beam. We can estimate to an order of magnitude that $10\,\mu$s timing resolution is possible, giving an approximate additional 10$^{-5}$ reduction in background from non-beam-related sources occurring during a beam-spill for a total rate of about $\sim10^{-6}\,$events/kg$\cdot$day, or $\sim6\times10^{-4}\,$events/m$^3\cdot$yr. This rate is much smaller than any of those predicted in Table~\ref{tab:model_output}.

\begin{figure*}[t!]
    \centering
    \includegraphics[scale=0.43]{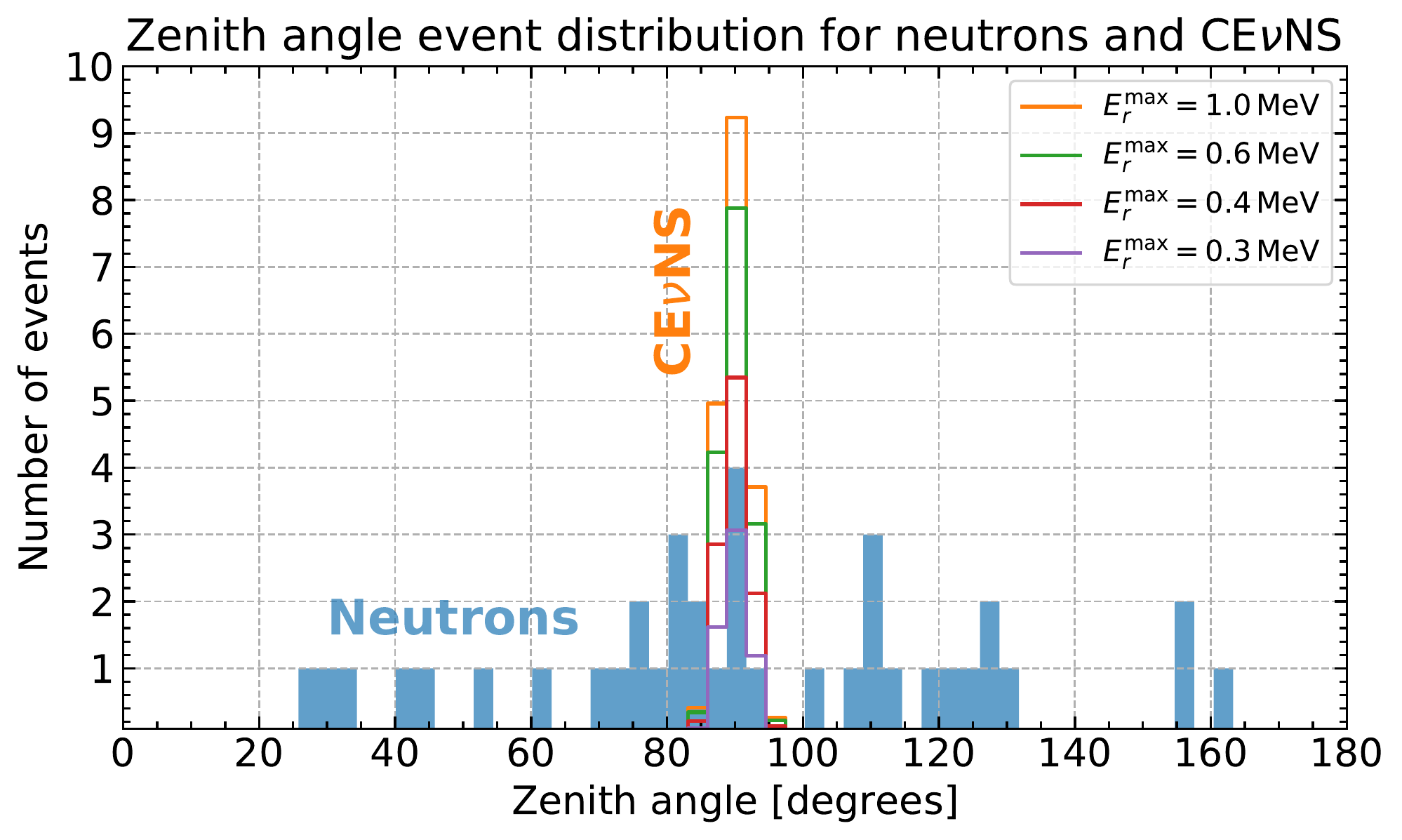}
    \includegraphics[scale=0.43]{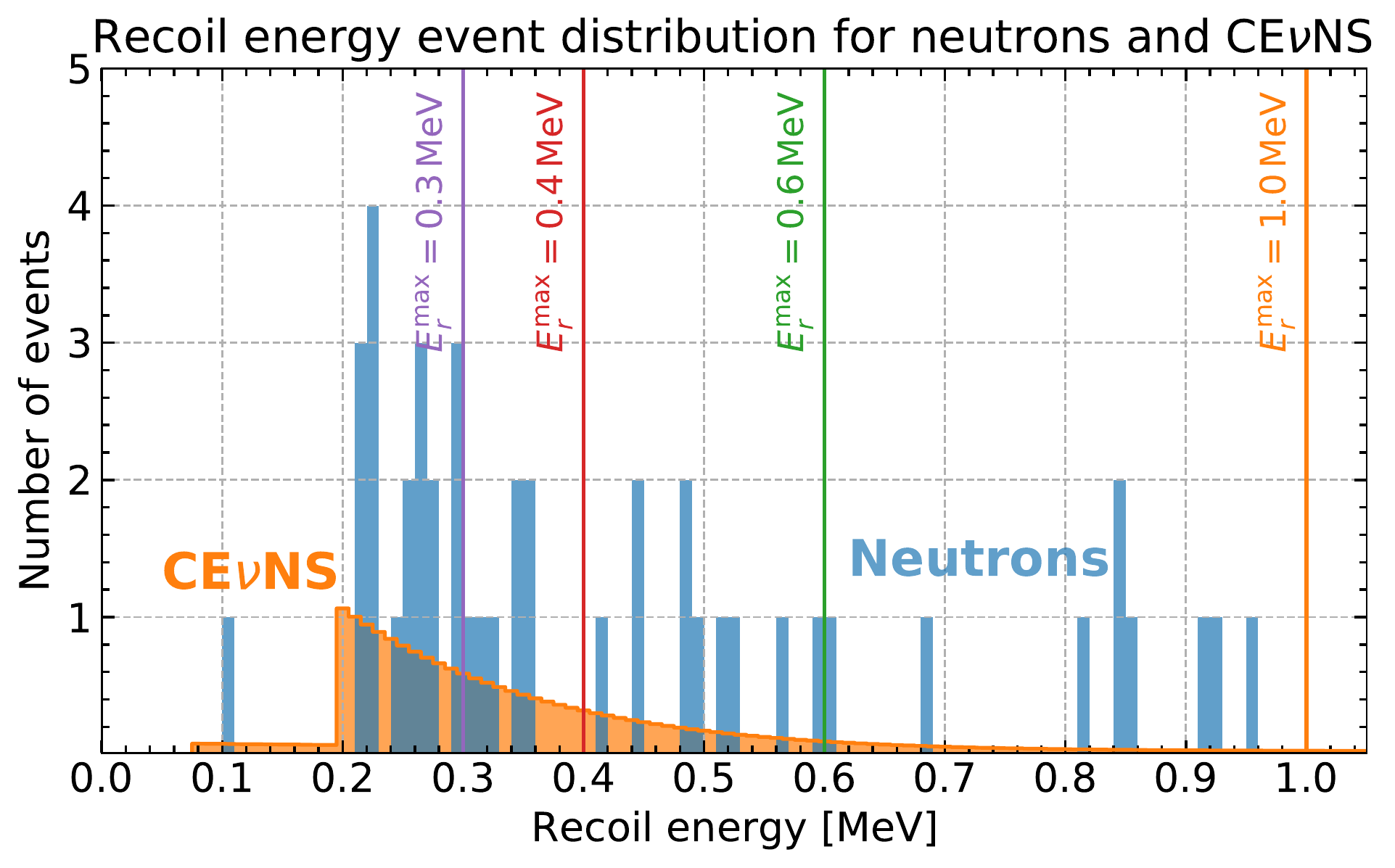}
    \caption{\textbf{Left graph}: Neutron and CE$\nu$NS zenith angle distribution as a function of zenith angle in degrees. The result has been derived assuming the NuMI LE neutrino flux, with parameters as specified in Table~\ref{tab:rock_components}. As expected, the CE$\nu$NS signal peaks at $90^\circ$ while the neutron-induce recoils have a much wider spread (see text in Sec.~\ref{sec:signal_backgrounds} for details). The histograms for different maximum recoil energies show that events pile up with increasing energy. \textbf{Right graph}: Neutron and CE$\nu$NS recoil energy spectra as a function of nuclear recoil energy. The result has been derived with the same assumptions that those used for the left graph. The different energy lines are correlated with the zenith angle histogram in the left graph and graphically indicate the number of events that for that energy have been piled up in the zenith angle distribution peak.}
    \label{fig:signal_background_neutrons_CEvNS}
\end{figure*}

\medskip

\section{Signal and rock neutrons backgrounds}
\label{sec:signal_backgrounds}

Our results demonstrate that un-vetoed, rock neutron backgrounds can be substantial, in particular for the NuMI HE mode and the DUNE on-axis configuration. Further discrimination of the CE$\nu$NS signal against this background would be helpful. To do so the directional capabilities of the detector can be employed. Information from the neutron and CE$\nu$NS zenith angle distribution spectra combined with their recoil energy spectra provide information that allows---in principle---efficient background discrimination. The CE$\nu$NS angular distribution is expected to peak in the direction perpendicular to the neutrino flux. This can be readily understood from the fact that the recoil (zenith) angle $\theta_r$ and recoil energy $E_r$ are related through~\citep{Abdullah:2020iiv}
\begin{equation}
    \label{eq:recoil_angle_recoil_energy_relation}
    \cos\theta_r = \sqrt{\frac{m_N\,E_r}{2}}
    \left(
    \frac{1}{E_\nu} + \frac{1}{m_N}
    \right)\,,
\end{equation}
where $\theta_r$ is the recoil angle relative to the direction of the neutrino, $m_N$ is the mass of the nucleus and $E_\nu$ is the energy of the neutrino. For the typical recoil energies ($<1\,$MeV), induced by a ``high-energy" neutrino beam ($\sim$GeV) as those we have consider in these simulations, lead to small $\cos\theta_r$. For CE$\nu$NS this translates into most events clustering at $90^\circ$, independent of the neutrino beam we choose. To exploit this fact the neutron zenith angle distribution has to be as well categorized. Its exact morphology, in contrast to the CE$\nu$NS signal, does depend on the neutrino flux and so for concreteness we have performed calculations for the NuMI LE mode. 

The left graph in Fig.~\ref{fig:signal_background_neutrons_CEvNS} shows the results for both spectra for a 10 m$^3$ year exposure. The neutron recoil angular distribution has a mild tendency to cluster at about $90^\circ$ due to a tendency of rock neutrons to preserve the forward direction of the beam. However, their spectrum has a much wider spread in comparison to that of neutrinos recoils. This result thus shows that with a reasonable angular resolution further discrimination ($\sim$10$^4$:1 altogether, scintillating veto plus angular cuts) of background events is possible. At $90^\circ$ the signal-to-background ratio is estimated by comparing the number of events at peak, is $\sim$2.5.

The recoil energy spectra provide, as well, useful discrimination power. To determine the degree to which by itself, or through its interplay with zenith angle spectral information this can be done, we have calculated the CE$\nu$NS recoil energy signal as well as neutron recoil energy spectra for the same neutrino flux configuration. Results are shown in the right graph in Fig.~\ref{fig:signal_background_neutrons_CEvNS}. The CE$\nu$NS signal spreads over a wider energy range (compared to its clustering at $90^\circ$) but does peak towards lower recoil energies. 
The rock neutron background peaks as well at low recoil energies, but in contrast to the CE$\nu$NS signal does populate the full energy range suggesting a different spectrum which could be exploited.

In addition some amount of C and S recoil discrimination is present. The difference in these spectra could be used to further discriminate the signals.
More work is needed to fully exploit the background rejection capability of these signatures.

Other backgrounds could be considered and studied. The decay-in-flight neutrino beam energies extend up to and even beyond $\sim10\,$GeV; thus, in addition to CE$\nu$NS, other higher-energy processes such as quasielastic, resonance, and deep-inelastic scattering will occur, see e.g.~\citep{Formaggio:2012cpf}. The cross sections for these higher-energy interactions (wherein the constituent nucleons become the system's dominant degrees of freedom) are sizable at higher $Q^2$. As discussed above for COUPP, these type of events occur at a rate of $0.35\,$events$/$kg$\cdot$day. The $\nu$BDX-DRIFT detector with a mass of 1.6 kg will see these events on the order of 1 per day. In terms of backgrounds to $\nu$BDX-DRIFT in searches for CE$\nu$NS and BSM nuclear recoils though, the large neutrino energies generally imply high particle multiplicities and are comparatively unique in their topologies. For instance, charged particles produced in conjunction with nuclear recoils can be rejected as signal events. As shown above, the scintillating veto is extremely effective at rejecting neutrals at these large energies. Additionally, as events like this will be present in the data, their characteristics can be measured and studied themselves, an interesting topic it's own right. 

\section{Conclusion}
\label{sec:conclusions}
In this paper we have studied rock neutron backgrounds in the $\nu$BDX-DRIFT detector. Rock neutrons are produced by the interaction of neutrinos with the rock surrounding the underground hall where the detector is deployed. End-state particles produced in these interactions come from a~\texttt{GENIE} Monte Carlo calculation which uses four possible neutrino fluxes (NuMI LE and HE modes and DUNE on-axis and off-axis 39 m configurations) interacting with the rock composed mainly of Oxygen, Silicon, Aluminum and Iron. The energy spectra of the final state particles produced in these interactions serve then as an input for a~\texttt{GEANT4} Monte Carlo simulation, which propagates these states throughout the rock and so allows the characterization of the neutrons emerging from the walls of the hall. These neutrons are then used to study the possible backgrounds to which the $\nu$BDX-DRIFT detector will be subject to while being operated at the FNAL. 

The simulation is bench-marked against the 2009 beam-tagged COUPP data, obtained by the COUPP collaboration during operation in the MINOS hall while the NuMI beamline was operated in the LE mode. Agreement between the simulated and actual data is found within ~30\%. After this validation, results for energy, zenith and azimuth spectra for the neutrons emitted by the walls are reported. These results, crucial for the determination of rock neutron backgrounds in the $\nu$BDX-DRIFT detector, are as well useful for future neutrino experiments at the FNAL. They add to undergoing efforts by the ANNIE collaboration, which aims to characterize neutron backgrounds at the FNAL.

With the ``morphology" of the emitted neutrons at hand, rock neutron backgrounds within the $\nu$BDX-DRIFT fiducial volume have been determined. By assuming the detector to be fully surrounded by a 75 cm thick BC-521 scintillator veto, for the four different neutrino flux configurations we have found that the DUNE off-axis 39 m provides the most background-suppressed experimental scenario. Rock neutron backgrounds gradually increase from the NuMI LE to the DUNE on-axis to the NuMI HE, with the latter being the configuration leading to the largest background. Detailed results have been reported in Table~\ref{tab:model_output}.

Finally we have discussed discrimination of rock neutron backgrounds against CE$\nu$NS signals. Using NuMI LE as a representative case, we have compared neutron and CE$\nu$NS zenith and recoil energy spectra. The results demonstrate that discrimination against rock neutron backgrounds is possible. Firstly, the CE$\nu$NS signal peaks at $90\degree$, in contrast to the neutron background that spreads more uniformly. At peak, the signal-to-background ratio has been roughly estimated to be $\sim 2.5$. Information from the recoil energy spectra shows that background-free energy windows exist, thus offering an experimental avenue for CE$\nu$NS measurements as well as for BSM searches.

\section*{Acknowledgements}
We thank Andrew Sonnenschein and Jeter Hall for useful conversations and providing us the the COUPP 2009 results upon which this work is based.  We thank Eric V\'azquez-J\'auregui for carefully reading the manuscript and useful comments, particularly concerning the actual geometry of the COUPP 2009 detector. D.A.S. is supported by ANID grant ``Fondecyt Regular" N$^\text{o}$ 1221445. BD and LES acknowledge support from DOE Grant de-sc0010813.

\bibliographystyle{apsrev4-1}
\bibliography{Backgrounds}

\begin{thebibliography}{47}%
\makeatletter
\providecommand \@ifxundefined [1]{%
 \@ifx{#1\undefined}
}%
\providecommand \@ifnum [1]{%
 \ifnum #1\expandafter \@firstoftwo
 \else \expandafter \@secondoftwo
 \fi
}%
\providecommand \@ifx [1]{%
 \ifx #1\expandafter \@firstoftwo
 \else \expandafter \@secondoftwo
 \fi
}%
\providecommand \natexlab [1]{#1}%
\providecommand \enquote  [1]{``#1''}%
\providecommand \bibnamefont  [1]{#1}%
\providecommand \bibfnamefont [1]{#1}%
\providecommand \citenamefont [1]{#1}%
\providecommand \href@noop [0]{\@secondoftwo}%
\providecommand \href [0]{\begingroup \@sanitize@url \@href}%
\providecommand \@href[1]{\@@startlink{#1}\@@href}%
\providecommand \@@href[1]{\endgroup#1\@@endlink}%
\providecommand \@sanitize@url [0]{\catcode `\\12\catcode `\$12\catcode
  `\&12\catcode `\#12\catcode `\^12\catcode `\_12\catcode `\%12\relax}%
\providecommand \@@startlink[1]{}%
\providecommand \@@endlink[0]{}%
\providecommand \url  [0]{\begingroup\@sanitize@url \@url }%
\providecommand \@url [1]{\endgroup\@href {#1}{\urlprefix }}%
\providecommand \urlprefix  [0]{URL }%
\providecommand \Eprint [0]{\href }%
\providecommand \doibase [0]{http://dx.doi.org/}%
\providecommand \selectlanguage [0]{\@gobble}%
\providecommand \bibinfo  [0]{\@secondoftwo}%
\providecommand \bibfield  [0]{\@secondoftwo}%
\providecommand \translation [1]{[#1]}%
\providecommand \BibitemOpen [0]{}%
\providecommand \bibitemStop [0]{}%
\providecommand \bibitemNoStop [0]{.\EOS\space}%
\providecommand \EOS [0]{\spacefactor3000\relax}%
\providecommand \BibitemShut  [1]{\csname bibitem#1\endcsname}%
\let\auto@bib@innerbib\@empty
\bibitem [{\citenamefont {Freedman}(1974)}]{Freedman:1973yd}%
  \BibitemOpen
  \bibfield  {author} {\bibinfo {author} {\bibfnamefont {D.~Z.}\ \bibnamefont
  {Freedman}},\ }\href {\doibase 10.1103/PhysRevD.9.1389} {\bibfield  {journal}
  {\bibinfo  {journal} {Phys. Rev. D}\ }\textbf {\bibinfo {volume} {9}},\
  \bibinfo {pages} {1389} (\bibinfo {year} {1974})}\BibitemShut {NoStop}%
\bibitem [{\citenamefont {Freedman}\ \emph {et~al.}(1977)\citenamefont
  {Freedman}, \citenamefont {Schramm},\ and\ \citenamefont
  {Tubbs}}]{Freedman:1977xn}%
  \BibitemOpen
  \bibfield  {author} {\bibinfo {author} {\bibfnamefont {D.~Z.}\ \bibnamefont
  {Freedman}}, \bibinfo {author} {\bibfnamefont {D.~N.}\ \bibnamefont
  {Schramm}}, \ and\ \bibinfo {author} {\bibfnamefont {D.~L.}\ \bibnamefont
  {Tubbs}},\ }\href {\doibase 10.1146/annurev.ns.27.120177.001123} {\bibfield
  {journal} {\bibinfo  {journal} {Ann. Rev. Nucl. Part. Sci.}\ }\textbf
  {\bibinfo {volume} {27}},\ \bibinfo {pages} {167} (\bibinfo {year}
  {1977})}\BibitemShut {NoStop}%
\bibitem [{\citenamefont {Aristizabal~Sierra}\ \emph
  {et~al.}(2021)\citenamefont {Aristizabal~Sierra}, \citenamefont {Dutta},
  \citenamefont {Kim}, \citenamefont {Snowden-Ifft},\ and\ \citenamefont
  {Strigari}}]{PhysRevD.104.033004}%
  \BibitemOpen
  \bibfield  {author} {\bibinfo {author} {\bibfnamefont {D.}~\bibnamefont
  {Aristizabal~Sierra}}, \bibinfo {author} {\bibfnamefont {B.}~\bibnamefont
  {Dutta}}, \bibinfo {author} {\bibfnamefont {D.}~\bibnamefont {Kim}}, \bibinfo
  {author} {\bibfnamefont {D.}~\bibnamefont {Snowden-Ifft}}, \ and\ \bibinfo
  {author} {\bibfnamefont {L.~E.}\ \bibnamefont {Strigari}},\ }\href {\doibase
  10.1103/PhysRevD.104.033004} {\bibfield  {journal} {\bibinfo  {journal}
  {Phys. Rev. D}\ }\textbf {\bibinfo {volume} {104}},\ \bibinfo {pages}
  {033004} (\bibinfo {year} {2021})}\BibitemShut {NoStop}%
\bibitem [{\citenamefont {Abdullah}\ \emph {et~al.}(2020)\citenamefont
  {Abdullah}, \citenamefont {Aristizabal~Sierra}, \citenamefont {Dutta},\ and\
  \citenamefont {Strigari}}]{Abdullah:2020iiv}%
  \BibitemOpen
  \bibfield  {author} {\bibinfo {author} {\bibfnamefont {M.}~\bibnamefont
  {Abdullah}}, \bibinfo {author} {\bibfnamefont {D.}~\bibnamefont
  {Aristizabal~Sierra}}, \bibinfo {author} {\bibfnamefont {B.}~\bibnamefont
  {Dutta}}, \ and\ \bibinfo {author} {\bibfnamefont {L.~E.}\ \bibnamefont
  {Strigari}},\ }\href {\doibase 10.1103/PhysRevD.102.015009} {\bibfield
  {journal} {\bibinfo  {journal} {Phys. Rev. D}\ }\textbf {\bibinfo {volume}
  {102}},\ \bibinfo {pages} {015009} (\bibinfo {year} {2020})},\ \Eprint
  {http://arxiv.org/abs/2003.11510} {arXiv:2003.11510 [hep-ph]} \BibitemShut
  {NoStop}%
\bibitem [{\citenamefont {Strait}\ \emph {et~al.}(2016)\citenamefont {Strait}
  \emph {et~al.}}]{DUNE:2016evb}%
  \BibitemOpen
  \bibfield  {author} {\bibinfo {author} {\bibfnamefont {J.}~\bibnamefont
  {Strait}} \emph {et~al.} (\bibinfo {collaboration} {DUNE}),\ }\href@noop {}
  {\  (\bibinfo {year} {2016})},\ \Eprint {http://arxiv.org/abs/1601.05823}
  {arXiv:1601.05823 [physics.ins-det]} \BibitemShut {NoStop}%
\bibitem [{\citenamefont {Akimov}\ \emph {et~al.}(2017)\citenamefont {Akimov}
  \emph {et~al.}}]{COHERENT:2017ipa}%
  \BibitemOpen
  \bibfield  {author} {\bibinfo {author} {\bibfnamefont {D.}~\bibnamefont
  {Akimov}} \emph {et~al.} (\bibinfo {collaboration} {COHERENT}),\ }\href
  {\doibase 10.1126/science.aao0990} {\bibfield  {journal} {\bibinfo  {journal}
  {Science}\ }\textbf {\bibinfo {volume} {357}},\ \bibinfo {pages} {1123}
  (\bibinfo {year} {2017})},\ \Eprint {http://arxiv.org/abs/1708.01294}
  {arXiv:1708.01294 [nucl-ex]} \BibitemShut {NoStop}%
\bibitem [{\citenamefont {Akimov}\ \emph {et~al.}(2021)\citenamefont {Akimov}
  \emph {et~al.}}]{COHERENT:2020iec}%
  \BibitemOpen
  \bibfield  {author} {\bibinfo {author} {\bibfnamefont {D.}~\bibnamefont
  {Akimov}} \emph {et~al.} (\bibinfo {collaboration} {COHERENT}),\ }\href
  {\doibase 10.1103/PhysRevLett.126.012002} {\bibfield  {journal} {\bibinfo
  {journal} {Phys. Rev. Lett.}\ }\textbf {\bibinfo {volume} {126}},\ \bibinfo
  {pages} {012002} (\bibinfo {year} {2021})},\ \Eprint
  {http://arxiv.org/abs/2003.10630} {arXiv:2003.10630 [nucl-ex]} \BibitemShut
  {NoStop}%
\bibitem [{\citenamefont {Abdullah}\ \emph {et~al.}(2022)\citenamefont
  {Abdullah} \emph {et~al.}}]{Abdullah:2022zue}%
  \BibitemOpen
  \bibfield  {author} {\bibinfo {author} {\bibfnamefont {M.}~\bibnamefont
  {Abdullah}} \emph {et~al.},\ }\href@noop {} {\  (\bibinfo {year} {2022})},\
  \Eprint {http://arxiv.org/abs/2203.07361} {arXiv:2203.07361 [hep-ph]}
  \BibitemShut {NoStop}%
\bibitem [{\citenamefont {Aguilar-Arevalo}\ \emph {et~al.}(2019)\citenamefont
  {Aguilar-Arevalo} \emph {et~al.}}]{CONNIE:2019swq}%
  \BibitemOpen
  \bibfield  {author} {\bibinfo {author} {\bibfnamefont {A.}~\bibnamefont
  {Aguilar-Arevalo}} \emph {et~al.} (\bibinfo {collaboration} {CONNIE}),\
  }\href {\doibase 10.1103/PhysRevD.100.092005} {\bibfield  {journal} {\bibinfo
   {journal} {Phys. Rev. D}\ }\textbf {\bibinfo {volume} {100}},\ \bibinfo
  {pages} {092005} (\bibinfo {year} {2019})},\ \Eprint
  {http://arxiv.org/abs/1906.02200} {arXiv:1906.02200 [physics.ins-det]}
  \BibitemShut {NoStop}%
\bibitem [{\citenamefont {Agnolet}\ \emph {et~al.}(2017)\citenamefont {Agnolet}
  \emph {et~al.}}]{MINER:2016igy}%
  \BibitemOpen
  \bibfield  {author} {\bibinfo {author} {\bibfnamefont {G.}~\bibnamefont
  {Agnolet}} \emph {et~al.} (\bibinfo {collaboration} {MINER}),\ }\href
  {\doibase 10.1016/j.nima.2017.02.024} {\bibfield  {journal} {\bibinfo
  {journal} {Nucl. Instrum. Meth. A}\ }\textbf {\bibinfo {volume} {853}},\
  \bibinfo {pages} {53} (\bibinfo {year} {2017})},\ \Eprint
  {http://arxiv.org/abs/1609.02066} {arXiv:1609.02066 [physics.ins-det]}
  \BibitemShut {NoStop}%
\bibitem [{\citenamefont {Strauss}\ \emph {et~al.}(2017)\citenamefont {Strauss}
  \emph {et~al.}}]{Strauss:2017cuu}%
  \BibitemOpen
  \bibfield  {author} {\bibinfo {author} {\bibfnamefont {R.}~\bibnamefont
  {Strauss}} \emph {et~al.},\ }\href {\doibase 10.1140/epjc/s10052-017-5068-2}
  {\bibfield  {journal} {\bibinfo  {journal} {Eur. Phys. J. C}\ }\textbf
  {\bibinfo {volume} {77}},\ \bibinfo {pages} {506} (\bibinfo {year} {2017})},\
  \Eprint {http://arxiv.org/abs/1704.04320} {arXiv:1704.04320
  [physics.ins-det]} \BibitemShut {NoStop}%
\bibitem [{\citenamefont {Akimov}\ \emph {et~al.}(2020)\citenamefont {Akimov}
  \emph {et~al.}}]{RED-100:2019rpf}%
  \BibitemOpen
  \bibfield  {author} {\bibinfo {author} {\bibfnamefont {D.~Y.}\ \bibnamefont
  {Akimov}} \emph {et~al.} (\bibinfo {collaboration} {RED-100}),\ }\href
  {\doibase 10.1088/1748-0221/15/02/P02020} {\bibfield  {journal} {\bibinfo
  {journal} {JINST}\ }\textbf {\bibinfo {volume} {15}},\ \bibinfo {pages}
  {P02020} (\bibinfo {year} {2020})},\ \Eprint
  {http://arxiv.org/abs/1910.06190} {arXiv:1910.06190 [physics.ins-det]}
  \BibitemShut {NoStop}%
\bibitem [{\citenamefont {Colaresi}\ \emph {et~al.}(2022)\citenamefont
  {Colaresi}, \citenamefont {Collar}, \citenamefont {Hossbach}, \citenamefont
  {Lewis},\ and\ \citenamefont {Yocum}}]{Colaresi:2022obx}%
  \BibitemOpen
  \bibfield  {author} {\bibinfo {author} {\bibfnamefont {J.}~\bibnamefont
  {Colaresi}}, \bibinfo {author} {\bibfnamefont {J.~I.}\ \bibnamefont
  {Collar}}, \bibinfo {author} {\bibfnamefont {T.~W.}\ \bibnamefont
  {Hossbach}}, \bibinfo {author} {\bibfnamefont {C.~M.}\ \bibnamefont {Lewis}},
  \ and\ \bibinfo {author} {\bibfnamefont {K.~M.}\ \bibnamefont {Yocum}},\
  }\href@noop {} {\  (\bibinfo {year} {2022})},\ \Eprint
  {http://arxiv.org/abs/2202.09672} {arXiv:2202.09672 [hep-ex]} \BibitemShut
  {NoStop}%
\bibitem [{\citenamefont {Billard}\ \emph {et~al.}(2017)\citenamefont {Billard}
  \emph {et~al.}}]{Billard:2016giu}%
  \BibitemOpen
  \bibfield  {author} {\bibinfo {author} {\bibfnamefont {J.}~\bibnamefont
  {Billard}} \emph {et~al.},\ }\href {\doibase 10.1088/1361-6471/aa83d0}
  {\bibfield  {journal} {\bibinfo  {journal} {J. Phys. G}\ }\textbf {\bibinfo
  {volume} {44}},\ \bibinfo {pages} {105101} (\bibinfo {year} {2017})},\
  \Eprint {http://arxiv.org/abs/1612.09035} {arXiv:1612.09035
  [physics.ins-det]} \BibitemShut {NoStop}%
\bibitem [{\citenamefont {Bonet}\ \emph {et~al.}(2021)\citenamefont {Bonet}
  \emph {et~al.}}]{CONUS:2020skt}%
  \BibitemOpen
  \bibfield  {author} {\bibinfo {author} {\bibfnamefont {H.}~\bibnamefont
  {Bonet}} \emph {et~al.} (\bibinfo {collaboration} {CONUS}),\ }\href {\doibase
  10.1103/PhysRevLett.126.041804} {\bibfield  {journal} {\bibinfo  {journal}
  {Phys. Rev. Lett.}\ }\textbf {\bibinfo {volume} {126}},\ \bibinfo {pages}
  {041804} (\bibinfo {year} {2021})},\ \Eprint
  {http://arxiv.org/abs/2011.00210} {arXiv:2011.00210 [hep-ex]} \BibitemShut
  {NoStop}%
\bibitem [{\citenamefont {Akimov}\ \emph {et~al.}(2022)\citenamefont {Akimov}
  \emph {et~al.}}]{Akimov:2022oyb}%
  \BibitemOpen
  \bibfield  {author} {\bibinfo {author} {\bibfnamefont {D.}~\bibnamefont
  {Akimov}} \emph {et~al.},\ }in\ \href@noop {} {\emph {\bibinfo {booktitle}
  {{2022 Snowmass Summer Study}}}}\ (\bibinfo {year} {2022})\ \Eprint
  {http://arxiv.org/abs/2204.04575} {arXiv:2204.04575 [hep-ex]} \BibitemShut
  {NoStop}%
\bibitem [{\citenamefont {Garoby}\ \emph {et~al.}(2018)\citenamefont {Garoby}
  \emph {et~al.}}]{Garoby:2017vew}%
  \BibitemOpen
  \bibfield  {author} {\bibinfo {author} {\bibfnamefont {R.}~\bibnamefont
  {Garoby}} \emph {et~al.},\ }\href {\doibase 10.1088/1402-4896/aa9bff}
  {\bibfield  {journal} {\bibinfo  {journal} {Phys. Scripta}\ }\textbf
  {\bibinfo {volume} {93}},\ \bibinfo {pages} {014001} (\bibinfo {year}
  {2018})}\BibitemShut {NoStop}%
\bibitem [{\citenamefont {Andreopoulos}\ \emph {et~al.}(2010)\citenamefont
  {Andreopoulos} \emph {et~al.}}]{Andreopoulos:2009rq}%
  \BibitemOpen
  \bibfield  {author} {\bibinfo {author} {\bibfnamefont {C.}~\bibnamefont
  {Andreopoulos}} \emph {et~al.},\ }\href {\doibase 10.1016/j.nima.2009.12.009}
  {\bibfield  {journal} {\bibinfo  {journal} {Nucl. Instrum. Meth. A}\ }\textbf
  {\bibinfo {volume} {614}},\ \bibinfo {pages} {87} (\bibinfo {year} {2010})},\
  \Eprint {http://arxiv.org/abs/0905.2517} {arXiv:0905.2517 [hep-ph]}
  \BibitemShut {NoStop}%
\bibitem [{\citenamefont {Ambats}\ \emph {et~al.}(1998)\citenamefont {Ambats}
  \emph {et~al.}}]{MINOS:1998kez}%
  \BibitemOpen
  \bibfield  {author} {\bibinfo {author} {\bibfnamefont {I.}~\bibnamefont
  {Ambats}} \emph {et~al.} (\bibinfo {collaboration} {MINOS}),\ }\href
  {\doibase 10.2172/1861363} {\bibfield  {journal} {\bibinfo  {journal}
  {NUMI-L-337, FERMILAB-DESIGN-1998-02}\ } (\bibinfo {year} {1998}),\
  10.2172/1861363}\BibitemShut {NoStop}%
\bibitem [{\citenamefont {Agostinelli}\ \emph
  {et~al.}(2003{\natexlab{a}})\citenamefont {Agostinelli} \emph
  {et~al.}}]{GEANT4:2002zbu}%
  \BibitemOpen
  \bibfield  {author} {\bibinfo {author} {\bibfnamefont {S.}~\bibnamefont
  {Agostinelli}} \emph {et~al.} (\bibinfo {collaboration} {GEANT4}),\ }\href
  {\doibase 10.1016/S0168-9002(03)01368-8} {\bibfield  {journal} {\bibinfo
  {journal} {Nucl. Instrum. Meth. A}\ }\textbf {\bibinfo {volume} {506}},\
  \bibinfo {pages} {250} (\bibinfo {year} {2003}{\natexlab{a}})}\BibitemShut
  {NoStop}%
\bibitem [{\citenamefont {Adamson}\ \emph {et~al.}(2016)\citenamefont {Adamson}
  \emph {et~al.}}]{Adamson:2015dkw}%
  \BibitemOpen
  \bibfield  {author} {\bibinfo {author} {\bibfnamefont {P.}~\bibnamefont
  {Adamson}} \emph {et~al.},\ }\href {\doibase 10.1016/j.nima.2015.08.063}
  {\bibfield  {journal} {\bibinfo  {journal} {Nucl. Instrum. Meth. A}\ }\textbf
  {\bibinfo {volume} {806}},\ \bibinfo {pages} {279} (\bibinfo {year}
  {2016})},\ \Eprint {http://arxiv.org/abs/1507.06690} {arXiv:1507.06690
  [physics.acc-ph]} \BibitemShut {NoStop}%
\bibitem [{\citenamefont {Abi}\ \emph {et~al.}(2020)\citenamefont {Abi} \emph
  {et~al.}}]{DUNE:2020ypp}%
  \BibitemOpen
  \bibfield  {author} {\bibinfo {author} {\bibfnamefont {B.}~\bibnamefont
  {Abi}} \emph {et~al.} (\bibinfo {collaboration} {DUNE}),\ }\href@noop {} {\
  (\bibinfo {year} {2020})},\ \Eprint {http://arxiv.org/abs/2002.03005}
  {arXiv:2002.03005 [hep-ex]} \BibitemShut {NoStop}%
\bibitem [{\citenamefont {Anghel}\ \emph {et~al.}(2015)\citenamefont {Anghel}
  \emph {et~al.}}]{ANNIE:2015inw}%
  \BibitemOpen
  \bibfield  {author} {\bibinfo {author} {\bibfnamefont {I.}~\bibnamefont
  {Anghel}} \emph {et~al.} (\bibinfo {collaboration} {ANNIE}),\ }\href@noop {}
  {\  (\bibinfo {year} {2015})},\ \Eprint {http://arxiv.org/abs/1504.01480}
  {arXiv:1504.01480 [physics.ins-det]} \BibitemShut {NoStop}%
\bibitem [{\citenamefont {Papoulias}\ and\ \citenamefont
  {Kosmas}(2018)}]{Papoulias:2017qdn}%
  \BibitemOpen
  \bibfield  {author} {\bibinfo {author} {\bibfnamefont {D.~K.}\ \bibnamefont
  {Papoulias}}\ and\ \bibinfo {author} {\bibfnamefont {T.~S.}\ \bibnamefont
  {Kosmas}},\ }\href {\doibase 10.1103/PhysRevD.97.033003} {\bibfield
  {journal} {\bibinfo  {journal} {Phys. Rev. D}\ }\textbf {\bibinfo {volume}
  {97}},\ \bibinfo {pages} {033003} (\bibinfo {year} {2018})},\ \Eprint
  {http://arxiv.org/abs/1711.09773} {arXiv:1711.09773 [hep-ph]} \BibitemShut
  {NoStop}%
\bibitem [{\citenamefont {Aristizabal~Sierra}\ \emph
  {et~al.}(2019{\natexlab{a}})\citenamefont {Aristizabal~Sierra}, \citenamefont
  {De~Romeri},\ and\ \citenamefont {Rojas}}]{AristizabalSierra:2019ufd}%
  \BibitemOpen
  \bibfield  {author} {\bibinfo {author} {\bibfnamefont {D.}~\bibnamefont
  {Aristizabal~Sierra}}, \bibinfo {author} {\bibfnamefont {V.}~\bibnamefont
  {De~Romeri}}, \ and\ \bibinfo {author} {\bibfnamefont {N.}~\bibnamefont
  {Rojas}},\ }\href {\doibase 10.1007/JHEP09(2019)069} {\bibfield  {journal}
  {\bibinfo  {journal} {JHEP}\ }\textbf {\bibinfo {volume} {09}},\ \bibinfo
  {pages} {069} (\bibinfo {year} {2019}{\natexlab{a}})},\ \Eprint
  {http://arxiv.org/abs/1906.01156} {arXiv:1906.01156 [hep-ph]} \BibitemShut
  {NoStop}%
\bibitem [{\citenamefont {Aristizabal~Sierra}\ \emph
  {et~al.}(2019{\natexlab{b}})\citenamefont {Aristizabal~Sierra}, \citenamefont
  {Dutta}, \citenamefont {Liao},\ and\ \citenamefont
  {Strigari}}]{AristizabalSierra:2019ykk}%
  \BibitemOpen
  \bibfield  {author} {\bibinfo {author} {\bibfnamefont {D.}~\bibnamefont
  {Aristizabal~Sierra}}, \bibinfo {author} {\bibfnamefont {B.}~\bibnamefont
  {Dutta}}, \bibinfo {author} {\bibfnamefont {S.}~\bibnamefont {Liao}}, \ and\
  \bibinfo {author} {\bibfnamefont {L.~E.}\ \bibnamefont {Strigari}},\ }\href
  {\doibase 10.1007/JHEP12(2019)124} {\bibfield  {journal} {\bibinfo  {journal}
  {JHEP}\ }\textbf {\bibinfo {volume} {12}},\ \bibinfo {pages} {124} (\bibinfo
  {year} {2019}{\natexlab{b}})},\ \Eprint {http://arxiv.org/abs/1910.12437}
  {arXiv:1910.12437 [hep-ph]} \BibitemShut {NoStop}%
\bibitem [{\citenamefont {Aristizabal~Sierra}\ \emph
  {et~al.}(2022)\citenamefont {Aristizabal~Sierra}, \citenamefont {De~Romeri},\
  and\ \citenamefont {Papoulias}}]{AristizabalSierra:2022axl}%
  \BibitemOpen
  \bibfield  {author} {\bibinfo {author} {\bibfnamefont {D.}~\bibnamefont
  {Aristizabal~Sierra}}, \bibinfo {author} {\bibfnamefont {V.}~\bibnamefont
  {De~Romeri}}, \ and\ \bibinfo {author} {\bibfnamefont {D.~K.}\ \bibnamefont
  {Papoulias}},\ }\href@noop {} {\  (\bibinfo {year} {2022})},\ \Eprint
  {http://arxiv.org/abs/2203.02414} {arXiv:2203.02414 [hep-ph]} \BibitemShut
  {NoStop}%
\bibitem [{\citenamefont {Coloma}\ \emph {et~al.}(2020)\citenamefont {Coloma},
  \citenamefont {Esteban}, \citenamefont {Gonzalez-Garcia},\ and\ \citenamefont
  {Maltoni}}]{Coloma:2019mbs}%
  \BibitemOpen
  \bibfield  {author} {\bibinfo {author} {\bibfnamefont {P.}~\bibnamefont
  {Coloma}}, \bibinfo {author} {\bibfnamefont {I.}~\bibnamefont {Esteban}},
  \bibinfo {author} {\bibfnamefont {M.~C.}\ \bibnamefont {Gonzalez-Garcia}}, \
  and\ \bibinfo {author} {\bibfnamefont {M.}~\bibnamefont {Maltoni}},\ }\href
  {\doibase 10.1007/JHEP02(2020)023} {\bibfield  {journal} {\bibinfo  {journal}
  {JHEP}\ }\textbf {\bibinfo {volume} {02}},\ \bibinfo {pages} {023} (\bibinfo
  {year} {2020})},\ \bibinfo {note} {[Addendum: JHEP 12, 071 (2020)]},\ \Eprint
  {http://arxiv.org/abs/1911.09109} {arXiv:1911.09109 [hep-ph]} \BibitemShut
  {NoStop}%
\bibitem [{\citenamefont {Abrahamyan}\ \emph {et~al.}(2012)\citenamefont
  {Abrahamyan} \emph {et~al.}}]{Abrahamyan:2012gp}%
  \BibitemOpen
  \bibfield  {author} {\bibinfo {author} {\bibfnamefont {S.}~\bibnamefont
  {Abrahamyan}} \emph {et~al.},\ }\href {\doibase
  10.1103/PhysRevLett.108.112502} {\bibfield  {journal} {\bibinfo  {journal}
  {Phys. Rev. Lett.}\ }\textbf {\bibinfo {volume} {108}},\ \bibinfo {pages}
  {112502} (\bibinfo {year} {2012})},\ \Eprint {http://arxiv.org/abs/1201.2568}
  {arXiv:1201.2568 [nucl-ex]} \BibitemShut {NoStop}%
\bibitem [{\citenamefont {Horowitz}\ \emph {et~al.}(2014)\citenamefont
  {Horowitz}, \citenamefont {Kumar},\ and\ \citenamefont
  {Michaels}}]{Horowitz:2013wha}%
  \BibitemOpen
  \bibfield  {author} {\bibinfo {author} {\bibfnamefont {C.~J.}\ \bibnamefont
  {Horowitz}}, \bibinfo {author} {\bibfnamefont {K.~S.}\ \bibnamefont {Kumar}},
  \ and\ \bibinfo {author} {\bibfnamefont {R.}~\bibnamefont {Michaels}},\
  }\href {\doibase 10.1140/epja/i2014-14048-3} {\bibfield  {journal} {\bibinfo
  {journal} {Eur. Phys. J. A}\ }\textbf {\bibinfo {volume} {50}},\ \bibinfo
  {pages} {48} (\bibinfo {year} {2014})},\ \Eprint
  {http://arxiv.org/abs/1307.3572} {arXiv:1307.3572 [nucl-ex]} \BibitemShut
  {NoStop}%
\bibitem [{\citenamefont {Miranda}\ \emph {et~al.}(2020)\citenamefont
  {Miranda}, \citenamefont {Papoulias}, \citenamefont {Sanchez~Garcia},
  \citenamefont {Sanders}, \citenamefont {T\'ortola},\ and\ \citenamefont
  {Valle}}]{Miranda:2020tif}%
  \BibitemOpen
  \bibfield  {author} {\bibinfo {author} {\bibfnamefont {O.~G.}\ \bibnamefont
  {Miranda}}, \bibinfo {author} {\bibfnamefont {D.~K.}\ \bibnamefont
  {Papoulias}}, \bibinfo {author} {\bibfnamefont {G.}~\bibnamefont
  {Sanchez~Garcia}}, \bibinfo {author} {\bibfnamefont {O.}~\bibnamefont
  {Sanders}}, \bibinfo {author} {\bibfnamefont {M.}~\bibnamefont {T\'ortola}},
  \ and\ \bibinfo {author} {\bibfnamefont {J.~W.~F.}\ \bibnamefont {Valle}},\
  }\href {\doibase 10.1007/JHEP05(2020)130} {\bibfield  {journal} {\bibinfo
  {journal} {JHEP}\ }\textbf {\bibinfo {volume} {05}},\ \bibinfo {pages} {130}
  (\bibinfo {year} {2020})},\ \bibinfo {note} {[Erratum: JHEP 01, 067
  (2021)]},\ \Eprint {http://arxiv.org/abs/2003.12050} {arXiv:2003.12050
  [hep-ph]} \BibitemShut {NoStop}%
\bibitem [{\citenamefont {de~Gouvea}\ \emph {et~al.}(2020)\citenamefont
  {de~Gouvea}, \citenamefont {Machado}, \citenamefont {Perez-Gonzalez},\ and\
  \citenamefont {Tabrizi}}]{deGouvea:2019wav}%
  \BibitemOpen
  \bibfield  {author} {\bibinfo {author} {\bibfnamefont {A.}~\bibnamefont
  {de~Gouvea}}, \bibinfo {author} {\bibfnamefont {P.~A.~N.}\ \bibnamefont
  {Machado}}, \bibinfo {author} {\bibfnamefont {Y.~F.}\ \bibnamefont
  {Perez-Gonzalez}}, \ and\ \bibinfo {author} {\bibfnamefont {Z.}~\bibnamefont
  {Tabrizi}},\ }\href {\doibase 10.1103/PhysRevLett.125.051803} {\bibfield
  {journal} {\bibinfo  {journal} {Phys. Rev. Lett.}\ }\textbf {\bibinfo
  {volume} {125}},\ \bibinfo {pages} {051803} (\bibinfo {year} {2020})},\
  \Eprint {http://arxiv.org/abs/1912.06658} {arXiv:1912.06658 [hep-ph]}
  \BibitemShut {NoStop}%
\bibitem [{\citenamefont {Behnke}\ \emph {et~al.}(2011)\citenamefont {Behnke},
  \citenamefont {Behnke}, \citenamefont {Brice}, \citenamefont {Broemmelsiek},
  \citenamefont {Collar}, \citenamefont {Cooper}, \citenamefont {Crisler},
  \citenamefont {Dahl}, \citenamefont {Fustin}, \citenamefont {Hall},
  \citenamefont {Hinnefeld}, \citenamefont {Hu}, \citenamefont {Levine},
  \citenamefont {Ramberg}, \citenamefont {Shepherd}, \citenamefont
  {Sonnenschein}, \citenamefont {Szydagis},\ and\ \citenamefont
  {Collaboration}}]{COUPP2011}%
  \BibitemOpen
  \bibfield  {author} {\bibinfo {author} {\bibfnamefont {E.}~\bibnamefont
  {Behnke}}, \bibinfo {author} {\bibfnamefont {J.}~\bibnamefont {Behnke}},
  \bibinfo {author} {\bibfnamefont {S.~J.}\ \bibnamefont {Brice}}, \bibinfo
  {author} {\bibfnamefont {D.}~\bibnamefont {Broemmelsiek}}, \bibinfo {author}
  {\bibfnamefont {J.~I.}\ \bibnamefont {Collar}}, \bibinfo {author}
  {\bibfnamefont {P.~S.}\ \bibnamefont {Cooper}}, \bibinfo {author}
  {\bibfnamefont {M.}~\bibnamefont {Crisler}}, \bibinfo {author} {\bibfnamefont
  {C.~E.}\ \bibnamefont {Dahl}}, \bibinfo {author} {\bibfnamefont
  {D.}~\bibnamefont {Fustin}}, \bibinfo {author} {\bibfnamefont
  {J.}~\bibnamefont {Hall}}, \bibinfo {author} {\bibfnamefont {J.~H.}\
  \bibnamefont {Hinnefeld}}, \bibinfo {author} {\bibfnamefont {M.}~\bibnamefont
  {Hu}}, \bibinfo {author} {\bibfnamefont {I.}~\bibnamefont {Levine}}, \bibinfo
  {author} {\bibfnamefont {E.}~\bibnamefont {Ramberg}}, \bibinfo {author}
  {\bibfnamefont {T.}~\bibnamefont {Shepherd}}, \bibinfo {author}
  {\bibfnamefont {A.}~\bibnamefont {Sonnenschein}}, \bibinfo {author}
  {\bibfnamefont {M.}~\bibnamefont {Szydagis}}, \ and\ \bibinfo {author}
  {\bibfnamefont {C.}~\bibnamefont {Collaboration}},\ }\href {\doibase
  10.1103/PhysRevLett.106.021303} {\bibfield  {journal} {\bibinfo  {journal}
  {Phys. Rev. Lett.}\ }\textbf {\bibinfo {volume} {106}} (\bibinfo {year}
  {2011}),\ 10.1103/PhysRevLett.106.021303}\BibitemShut {NoStop}%
\bibitem [{\citenamefont {{COUPP Collaboration}}(2020)}]{Priv_Comm}%
  \BibitemOpen
  \bibfield  {author} {\bibinfo {author} {\bibnamefont {{COUPP
  Collaboration}}},\ }\href@noop {} {}\bibinfo {howpublished} {{Private
  Communication}} (\bibinfo {year} {2020})\BibitemShut {NoStop}%
\bibitem [{\citenamefont {Yuan}()}]{DataLogger}%
  \BibitemOpen
  \bibfield  {author} {\bibinfo {author} {\bibfnamefont {Z.}~\bibnamefont
  {Yuan}},\ }\href {https://www-bd.fnal.gov/D44/d44_ajax.html} {\enquote
  {\bibinfo {title} {D44 - data logger plotter},}\ }\BibitemShut {NoStop}%
\bibitem [{\citenamefont {Kopp}(2007)}]{Kopp:2007iq}%
  \BibitemOpen
  \bibfield  {author} {\bibinfo {author} {\bibfnamefont {S.~E.}\ \bibnamefont
  {Kopp}},\ }\href@noop {} {\  (\bibinfo {year} {2007})},\ \Eprint
  {http://arxiv.org/abs/0709.2737} {arXiv:0709.2737 [hep-ex]} \BibitemShut
  {NoStop}%
\bibitem [{\citenamefont {Collaboration}()}]{MinosLogs}%
  \BibitemOpen
  \bibfield  {author} {\bibinfo {author} {\bibnamefont {Collaboration}}
  (\bibinfo {collaboration} {MINOS}),\ }\href
  {https://www-users.cse.umn.edu/~straitm/minos/MINOS_runs} {\enquote {\bibinfo
  {title} {Minos run logs},}\ }\BibitemShut {NoStop}%
\bibitem [{\citenamefont {Aliaga~Soplin}(2016)}]{AliagaSoplin:2016shs}%
  \BibitemOpen
  \bibfield  {author} {\bibinfo {author} {\bibfnamefont {L.}~\bibnamefont
  {Aliaga~Soplin}},\ }\emph {\bibinfo {title} {{Neutrino Flux Prediction for
  the NuMI Beamline}}},\ \href {\doibase 10.2172/1250884} {Ph.D. thesis},\
  \bibinfo  {school} {William-Mary Coll.} (\bibinfo {year} {2016})\BibitemShut
  {NoStop}%
\bibitem [{\citenamefont {Andreopoulos}\ \emph {et~al.}(2015)\citenamefont
  {Andreopoulos}, \citenamefont {Barry}, \citenamefont {Dytman}, \citenamefont
  {Gallagher}, \citenamefont {Golan}, \citenamefont {Hatcher}, \citenamefont
  {Perdue},\ and\ \citenamefont {Yarba}}]{Andreopoulos:2015wxa}%
  \BibitemOpen
  \bibfield  {author} {\bibinfo {author} {\bibfnamefont {C.}~\bibnamefont
  {Andreopoulos}}, \bibinfo {author} {\bibfnamefont {C.}~\bibnamefont {Barry}},
  \bibinfo {author} {\bibfnamefont {S.}~\bibnamefont {Dytman}}, \bibinfo
  {author} {\bibfnamefont {H.}~\bibnamefont {Gallagher}}, \bibinfo {author}
  {\bibfnamefont {T.}~\bibnamefont {Golan}}, \bibinfo {author} {\bibfnamefont
  {R.}~\bibnamefont {Hatcher}}, \bibinfo {author} {\bibfnamefont
  {G.}~\bibnamefont {Perdue}}, \ and\ \bibinfo {author} {\bibfnamefont
  {J.}~\bibnamefont {Yarba}},\ }\href@noop {} {\  (\bibinfo {year} {2015})},\
  \Eprint {http://arxiv.org/abs/1510.05494} {arXiv:1510.05494 [hep-ph]}
  \BibitemShut {NoStop}%
\bibitem [{\citenamefont {Niewczas}\ and\ \citenamefont
  {Sobczyk}(2019)}]{Niewczas:2019fro}%
  \BibitemOpen
  \bibfield  {author} {\bibinfo {author} {\bibfnamefont {K.}~\bibnamefont
  {Niewczas}}\ and\ \bibinfo {author} {\bibfnamefont {J.~T.}\ \bibnamefont
  {Sobczyk}},\ }\href {\doibase 10.1103/PhysRevC.100.015505} {\bibfield
  {journal} {\bibinfo  {journal} {Phys. Rev. C}\ }\textbf {\bibinfo {volume}
  {100}},\ \bibinfo {pages} {015505} (\bibinfo {year} {2019})},\ \Eprint
  {http://arxiv.org/abs/1902.05618} {arXiv:1902.05618 [hep-ex]} \BibitemShut
  {NoStop}%
\bibitem [{\citenamefont {Golan}\ \emph {et~al.}(2012)\citenamefont {Golan},
  \citenamefont {Juszczak},\ and\ \citenamefont {Sobczyk}}]{Golan:2012wx}%
  \BibitemOpen
  \bibfield  {author} {\bibinfo {author} {\bibfnamefont {T.}~\bibnamefont
  {Golan}}, \bibinfo {author} {\bibfnamefont {C.}~\bibnamefont {Juszczak}}, \
  and\ \bibinfo {author} {\bibfnamefont {J.~T.}\ \bibnamefont {Sobczyk}},\
  }\href {\doibase 10.1103/PhysRevC.86.015505} {\bibfield  {journal} {\bibinfo
  {journal} {Phys. Rev. C}\ }\textbf {\bibinfo {volume} {86}},\ \bibinfo
  {pages} {015505} (\bibinfo {year} {2012})},\ \Eprint
  {http://arxiv.org/abs/1202.4197} {arXiv:1202.4197 [nucl-th]} \BibitemShut
  {NoStop}%
\bibitem [{\citenamefont {Agostinelli}\ \emph
  {et~al.}(2003{\natexlab{b}})\citenamefont {Agostinelli} \emph
  {et~al.}}]{AGOSTINELLI2003250}%
  \BibitemOpen
  \bibfield  {author} {\bibinfo {author} {\bibfnamefont {S.}~\bibnamefont
  {Agostinelli}} \emph {et~al.} (\bibinfo {collaboration} {GEANT4}),\ }\href
  {\doibase 10.1016/S0168-9002(03)01368-8} {\bibfield  {journal} {\bibinfo
  {journal} {Nucl. Instrum. Meth. A}\ }\textbf {\bibinfo {volume} {506}},\
  \bibinfo {pages} {250} (\bibinfo {year} {2003}{\natexlab{b}})}\BibitemShut
  {NoStop}%
\bibitem [{\citenamefont {Behnke}\ \emph {et~al.}(2013)\citenamefont {Behnke},
  \citenamefont {Benjamin}, \citenamefont {Brice}, \citenamefont
  {Broemmelsiek}, \citenamefont {Collar}, \citenamefont {Cooper}, \citenamefont
  {Crisler}, \citenamefont {Dahl}, \citenamefont {Fustin}, \citenamefont
  {Hall}, \citenamefont {Harnish}, \citenamefont {Levine}, \citenamefont
  {Lippincott}, \citenamefont {Moan}, \citenamefont {Nania}, \citenamefont
  {Neilson}, \citenamefont {Ramberg}, \citenamefont {Robinson}, \citenamefont
  {Ruschman}, \citenamefont {Sonnenschein}, \citenamefont {Vazquez-Jauregui},
  \citenamefont {Rivera}, \citenamefont {Uplegger},\ and\ \citenamefont
  {Collaboration}}]{BubbleNucleation}%
  \BibitemOpen
  \bibfield  {author} {\bibinfo {author} {\bibfnamefont {E.}~\bibnamefont
  {Behnke}}, \bibinfo {author} {\bibfnamefont {T.}~\bibnamefont {Benjamin}},
  \bibinfo {author} {\bibfnamefont {S.~J.}\ \bibnamefont {Brice}}, \bibinfo
  {author} {\bibfnamefont {D.}~\bibnamefont {Broemmelsiek}}, \bibinfo {author}
  {\bibfnamefont {J.~I.}\ \bibnamefont {Collar}}, \bibinfo {author}
  {\bibfnamefont {P.~S.}\ \bibnamefont {Cooper}}, \bibinfo {author}
  {\bibfnamefont {M.}~\bibnamefont {Crisler}}, \bibinfo {author} {\bibfnamefont
  {C.~E.}\ \bibnamefont {Dahl}}, \bibinfo {author} {\bibfnamefont
  {D.}~\bibnamefont {Fustin}}, \bibinfo {author} {\bibfnamefont
  {J.}~\bibnamefont {Hall}}, \bibinfo {author} {\bibfnamefont {C.}~\bibnamefont
  {Harnish}}, \bibinfo {author} {\bibfnamefont {I.}~\bibnamefont {Levine}},
  \bibinfo {author} {\bibfnamefont {W.~H.}\ \bibnamefont {Lippincott}},
  \bibinfo {author} {\bibfnamefont {T.}~\bibnamefont {Moan}}, \bibinfo {author}
  {\bibfnamefont {T.}~\bibnamefont {Nania}}, \bibinfo {author} {\bibfnamefont
  {R.}~\bibnamefont {Neilson}}, \bibinfo {author} {\bibfnamefont
  {E.}~\bibnamefont {Ramberg}}, \bibinfo {author} {\bibfnamefont {A.~E.}\
  \bibnamefont {Robinson}}, \bibinfo {author} {\bibfnamefont {M.}~\bibnamefont
  {Ruschman}}, \bibinfo {author} {\bibfnamefont {A.}~\bibnamefont
  {Sonnenschein}}, \bibinfo {author} {\bibfnamefont {E.}~\bibnamefont
  {Vazquez-Jauregui}}, \bibinfo {author} {\bibfnamefont {R.~A.}\ \bibnamefont
  {Rivera}}, \bibinfo {author} {\bibfnamefont {L.}~\bibnamefont {Uplegger}}, \
  and\ \bibinfo {author} {\bibfnamefont {C.}~\bibnamefont {Collaboration}},\
  }\href {\doibase 10.1103/PhysRevD.88.021101} {\bibfield  {journal} {\bibinfo
  {journal} {Phys. Rev. D}\ }\textbf {\bibinfo {volume} {88}} (\bibinfo {year}
  {2013}),\ 10.1103/PhysRevD.88.021101}\BibitemShut {NoStop}%
\bibitem [{\citenamefont {Gardiner}(2021{\natexlab{a}})}]{Gardiner:2020ulp}%
  \BibitemOpen
  \bibfield  {author} {\bibinfo {author} {\bibfnamefont {S.}~\bibnamefont
  {Gardiner}},\ }\href {\doibase 10.1103/PhysRevC.103.044604} {\bibfield
  {journal} {\bibinfo  {journal} {Phys. Rev. C}\ }\textbf {\bibinfo {volume}
  {103}},\ \bibinfo {pages} {044604} (\bibinfo {year} {2021}{\natexlab{a}})},\
  \Eprint {http://arxiv.org/abs/2010.02393} {arXiv:2010.02393 [nucl-th]}
  \BibitemShut {NoStop}%
\bibitem [{\citenamefont {Gardiner}(2021{\natexlab{b}})}]{Gardiner:2021qfr}%
  \BibitemOpen
  \bibfield  {author} {\bibinfo {author} {\bibfnamefont {S.}~\bibnamefont
  {Gardiner}},\ }\href {\doibase 10.1016/j.cpc.2021.108123} {\bibfield
  {journal} {\bibinfo  {journal} {Comput. Phys. Commun.}\ }\textbf {\bibinfo
  {volume} {269}},\ \bibinfo {pages} {108123} (\bibinfo {year}
  {2021}{\natexlab{b}})},\ \Eprint {http://arxiv.org/abs/2101.11867}
  {arXiv:2101.11867 [nucl-th]} \BibitemShut {NoStop}%
\bibitem [{\citenamefont {Burgos}\ \emph {et~al.}(2007)\citenamefont {Burgos},
  \citenamefont {Forbes}, \citenamefont {Ghag}, \citenamefont {Gold},
  \citenamefont {Kudryavtsev}, \citenamefont {Lawson}, \citenamefont {Loomba},
  \citenamefont {Majewski}, \citenamefont {Muna}, \citenamefont {Murphy},
  \citenamefont {Nicklin}, \citenamefont {Paling}, \citenamefont {Petkov},
  \citenamefont {Plank}, \citenamefont {Robinson}, \citenamefont {Sanghi},
  \citenamefont {Smith}, \citenamefont {Snowden-Ifft}, \citenamefont {Spooner},
  \citenamefont {Sumner}, \citenamefont {Turk},\ and\ \citenamefont
  {Tziaferi}}]{AstroPle.28.4.409}%
  \BibitemOpen
  \bibfield  {author} {\bibinfo {author} {\bibfnamefont {S.}~\bibnamefont
  {Burgos}}, \bibinfo {author} {\bibfnamefont {J.}~\bibnamefont {Forbes}},
  \bibinfo {author} {\bibfnamefont {C.}~\bibnamefont {Ghag}}, \bibinfo {author}
  {\bibfnamefont {M.}~\bibnamefont {Gold}}, \bibinfo {author} {\bibfnamefont
  {V.~A.}\ \bibnamefont {Kudryavtsev}}, \bibinfo {author} {\bibfnamefont
  {T.~B.}\ \bibnamefont {Lawson}}, \bibinfo {author} {\bibfnamefont
  {D.}~\bibnamefont {Loomba}}, \bibinfo {author} {\bibfnamefont
  {P.}~\bibnamefont {Majewski}}, \bibinfo {author} {\bibfnamefont
  {D.}~\bibnamefont {Muna}}, \bibinfo {author} {\bibfnamefont {A.~S.}\
  \bibnamefont {Murphy}}, \bibinfo {author} {\bibfnamefont {G.~G.}\
  \bibnamefont {Nicklin}}, \bibinfo {author} {\bibfnamefont {S.~M.}\
  \bibnamefont {Paling}}, \bibinfo {author} {\bibfnamefont {A.}~\bibnamefont
  {Petkov}}, \bibinfo {author} {\bibfnamefont {S.~J.~S.}\ \bibnamefont
  {Plank}}, \bibinfo {author} {\bibfnamefont {M.}~\bibnamefont {Robinson}},
  \bibinfo {author} {\bibfnamefont {N.}~\bibnamefont {Sanghi}}, \bibinfo
  {author} {\bibfnamefont {N.~J.~T.}\ \bibnamefont {Smith}}, \bibinfo {author}
  {\bibfnamefont {D.~P.}\ \bibnamefont {Snowden-Ifft}}, \bibinfo {author}
  {\bibfnamefont {N.~J.~C.}\ \bibnamefont {Spooner}}, \bibinfo {author}
  {\bibfnamefont {T.~J.}\ \bibnamefont {Sumner}}, \bibinfo {author}
  {\bibfnamefont {J.}~\bibnamefont {Turk}}, \ and\ \bibinfo {author}
  {\bibfnamefont {E.}~\bibnamefont {Tziaferi}},\ }\href {\doibase
  10.1016/j.astropartphys.2007.08.007} {\bibfield  {journal} {\bibinfo
  {journal} {ASTROPARTICLE PHYSICS}\ }\textbf {\bibinfo {volume} {28}},\
  \bibinfo {pages} {409} (\bibinfo {year} {2007})}\BibitemShut {NoStop}%
\bibitem [{\citenamefont {Formaggio}\ and\ \citenamefont
  {Zeller}(2012)}]{Formaggio:2012cpf}%
  \BibitemOpen
  \bibfield  {author} {\bibinfo {author} {\bibfnamefont {J.~A.}\ \bibnamefont
  {Formaggio}}\ and\ \bibinfo {author} {\bibfnamefont {G.~P.}\ \bibnamefont
  {Zeller}},\ }\href {\doibase 10.1103/RevModPhys.84.1307} {\bibfield
  {journal} {\bibinfo  {journal} {Rev. Mod. Phys.}\ }\textbf {\bibinfo {volume}
  {84}},\ \bibinfo {pages} {1307} (\bibinfo {year} {2012})},\ \Eprint
  {http://arxiv.org/abs/1305.7513} {arXiv:1305.7513 [hep-ex]} \BibitemShut
  {NoStop}%
\end{thebibliography}%
\end{document}